\def\b{\begin{equation}}
\def\e{\end{equation}}

\documentstyle[eqsecnum,prd,aps,psfig]{revtex}

\begin{document}

\title{Particle production and complex path analysis} 
\author{K.~Srinivasan\thanks{Electronic address:~srini@iucaa.ernet.in},  
T.~Padmanabhan\thanks{Electronic address:~paddy@iucaa.ernet.in}}
\address{IUCAA, Post Bag 4, Ganeshkhind, Pune 411 007, INDIA.}

\maketitle

\begin{abstract}
This paper discusses particle production in Schwarzchild-like 
spacetimes and in an uniform electric field. 
Both problems are approached using the method of complex 
path analysis which is used to describe tunnelling 
processes in semi-classical quantum mechanics.   
 
Particle production in Schwarzchild-like spacetimes with a 
horizon is obtained  here by a new and simple semi-classical 
method based on the method of complex paths.  
Hawking radiation is obtained in the $(t,r)$ co-ordinate 
system of the  standard Schwarzchild metric {\it without} 
requiring the Kruskal extension. 
The co-ordinate singularity present at the horizon 
manifests itself as a singularity in the expression 
for the semi-classical propagator for a scalar field. 
We give a prescription whereby this singularity is 
regularised with Hawking's result being recovered. 
The equation satisfied by a scalar field is also 
reduced to solving a one-dimensional effective 
Schr\"odinger equation with a potential $(-1/x^2)$ 
near the horizon. 
Constructing the action for a fictitious non-relativistic 
particle moving in this potential and applying the above 
mentioned prescription, one again recovers Hawking radiation.   

In the case of the electric field, standard quantum field 
theoretic methods can be used to obtain particle production 
in a purely time-dependent gauge. 
In a purely space-dependent gauge, however, the 
tunnelling interpretation has to be  resorted to in 
order to recover the previous result. 
We attempt, in this paper, to provide a tunnelling 
description using the formal method of complex paths for 
both the time and space dependent gauges. 
The usefulness of such a common description becomes evident 
when `mixed' gauges, which are functions of both space and 
time variables, are analysed.  
We report, in this paper, certain mixed gauges which have 
the interesting property that mode functions in these gauges 
are found to be a combination of {\it elementary} functions 
unlike the standard modes which are transcendental 
parabolic cylinder functions.   

Finally, we present an attempt to interpret particle 
production by the electric field as a tunnelling process 
between the two sectors of the Rindler spacetime. 
\end{abstract}

\reversemarginpar 
\draft

\section{Introduction and Summary}
In this paper, we present a critical review of particle production 
in Schwarzchild-like spacetimes and in an uniform electric field in 
Minkowski spacetime.   
We approach both problems by the method of complex paths 
discussed by Landau in~\cite{landau3} where it is used to 
describe tunnelling processes in non-relativistic 
semi-classical quantum mechanics.  
This powerful technique will be used as the basis to provide 
a new method of recovering Hawking radiation in the 
usual Schwarzchild co-ordinates without requiring 
the Kruskal extension. 
It will also be used to describe particle production in 
an electric field in different non-trivial gauges and to 
link particle production in an electric field to tunnelling 
processes occuring in the Rindler spacetime.

The Schwarzchild-like spacetimes we consider here are the usual 
black hole, the Rindler and the de-Sitter spacetimes. 
In the standard black-hole spacetime, particle production was 
obtained by Hartle and Hawking~\cite{hawking76} using 
semi-classical analysis.
In this method, the semi-classical propagator for a scalar 
field propagating in the Schwarzchild spacetime is analytically 
continued in the time variable $t$ to complex values.  
This analytic continuation gives the result that the probability of 
emission of particles from the past horizon is not the same 
as the probability of absorption into the future horizon. 
The ratio between these probabilities is of the form
\begin{equation}
P[{\rm emission}] = P[{\rm absorption}] e^{-\beta E}
\end{equation}
where $E$ is the energy of the particles and $\beta = 1/8\pi M$ 
is the standard Hawking temperature. 
The above relation is interpreted to be equivalent to a thermal 
distribution of particles in analogy with that observed in any 
system interacting with black body radiation. 
In the latter case,  the probability of emission of radiation by 
the system is related to the probability of absorption by the system 
by a similar relation as given above. 
In Hawking's derivation, the Kruskal extension is of vital 
importance in obtaining the thermal spectrum. 
\par
In this  paper, we propose an alternate derivation of Hawking 
radiation {\it without using the Kruskal extension}. 
Our motivation for using the standard Schwarzchild co-ordinates 
rather than the Kruskal system are the following: 
(1) The Schwarzchild spacetime is a static spacetime. 
It contains a global Killing vector $\xi^a$ and the 
symmetry generated by this vector is respected in the 
Schwarzchild co-ordinates $(t,r)$.  
That is, the surfaces $t=constant$ where $t$ is the 
Schwarzchild `time' variable (this Killing vector is 
space-like in the region interior to the horizon but it 
is still a symmetry of the system) has the same structure 
for all $t$; $g_{ab}$ is independent of $t$.  
(2) The surface area of spheres of constant `radial' 
co-ordinate $r$ happen to be $4\pi r^2$, which is that 
of a sphere in flat spacetime, and hence these surfaces 
can be used to measure $r$. 
In contrast, the Kruskal system is an explicitly 
{\it time-dependent} spacetime since metric components 
now depend on the Kruskal time co-ordinate. 
It does not possess a global Killing vector associated 
with surfaces $u=constant$ where $u$ is the Kruskal `time' 
co-ordinate. 
Indeed, the presence and structure of such a global 
Killing vector is not immediately apparent from the form 
of the metric in these co-ordinates.  
Surfaces of constant `radial' co-ordinate do not possess 
the same property that corresponding surfaces in the 
Schwarzchild spacetime do. 
Further, the sectors which contain the past horizon and the 
time-reversed copy of the usual Schwarzchild 
spacetime are unphysical.  
In any realistic collapse scenario, these sectors cannot exist. 
The non-static nature of the Kruskal manifold imply that 
the `in' and `out' vacua cannot be the same and explicit 
particle production can take place.     
\par 
The difficulty, of course, is that the standard 
Schwarzchild co-ordinates possess a co-ordinate 
singularity at the horizon. 
As we shall see, this bad behaviour of the co-ordinates 
appears as a singularity in the expression for the 
semi-classical propagator near the horizon and we 
have to provide a specific prescription to bypass it.   
{\it This prescription gives the same result as that 
obtained by Hawking and can be used in all spacetimes 
with a Schwarzchild-like metric}.  
Note that the method described above is fundamentally 
different from the usual method of calculating the Bogolubov 
coefficients for an eternal black hole given, for example, 
in Ref.~\cite{bandd82}. 
In Ref.~\cite{bandd82}, appropriate linear combinations of 
the Schwarzchild mode functions that are analytic on the 
full Kruskal manifold (except at the past and future 
singularites at $r=0$) are constructed.  
The scalar field is then expanded in terms of these modes 
with the vacuum state being the Kruskal vacuum.  
Such an expansion provides the appropriate connnection 
between the annihilation and creation operators for the 
scalar field in the Schwarzchild and Kruskal 
co-ordinate systems. 
Then, for a given time like observer in either the left or 
right Schwarzchild sector, whose vacuum state is the 
Schwarzchild vacuum, the number operator is easily 
calculated and is found to correspond to a thermal spectrum.
In our method, however, the action functional is constructed 
using the Hamilton-Jacobi method in the appropriate co-ordinates. 
The singularity in the action caused by the singular 
behaviour of the co-ordinate system (in the {\it unextended} 
Schwarzchild co-ordinates) at the horizon is handled by 
the prescription to obtain particle production. 
We also reduce the problem of the massive scalar field 
propagating in such a spacetime to an effective Schr\"odinger 
equation which has a singular effective potential near the horizon. 
In this case, the singular behaviour of the co-ordinates 
manifests itself as a singularity in the potential. 
The non-relativistic action for a particle moving under the 
influence of such a potential is constructed and the same 
prescription is used to bypass this singularity. 
We again recover Hawking radiation.   
\par
We next consider the problem of a scalar field propagating 
in flat spacetime in an uniform electric field background.
The total particle production due to the presence of the electric 
field upto the one-loop approximation is correctly calculated by 
the gauge invariant method proposed by Schwinger~\cite{schwinger}. 
The same problem can be reduced, in a time dependent gauge, to an 
equivalent Schr\"odinger equation with an inverted harmonic 
oscillator potential.
Such an equation can be solved exactly using the standard flat 
spacetime quantum field theoretic methods. 
Since the problem is  explicitly time dependent, the vacuum state 
at $t \to -\infty$ and at $t \to \infty$ are not the same. 
The Bogoliubov coefficients between the ``in'' and ``out'' vacua 
are easily calculated and the total particle production turns out 
to be the same as that calculated by the Schwinger method.  
However, if a space dependent gauge is used to describe the same field, 
the vacuum state of the field remains the same for all time and hence 
no particle production can take place.  
To recover the standard result, it is usual to resort to the tunnelling interpretation. This interpretation is useful since it provides a 
dynamical picture of particle production (see, for example \cite{brout95}) 
and is the only way in which the standard gauge invariant result 
can be recovered in a time independent gauge.
\par
In this paper, we attempt a tunnelling description based on the 
method of complex paths for both the time-dependent and 
time-independent gauges. 
This method is used to calculate the transmission and 
reflection coefficients (or the tunnelling coefficients) 
for the equivalent quantum mechanical problem.  
Then,  an interpretation of these coefficients, in order to 
explicitly obtain the standard gauge invariant result is provided.
However, the usefulness of the tunnelling description is 
seen when simple `mixed' gauges, which are functions of 
both space and time, are considered.  
In some gauges, the scalar wave equation can be reduced to 
solving effective Schr\"odinger equations in suitable new 
variables which are combinations of the usual 
spacetime co-ordinates. 
By applying the method of complex paths to the these 
Schr\"odinger equations, the tunnelling coefficients are seen 
to match those obtained either for the purely time or 
purely space dependent gauges. 
The gauge invariant result is now recovered using the 
appropriate interpretation needed to identify the 
Bogolubov coefficients. 
In certain other gauges, the mathematics becomes simpler 
and the field equation can be reduced to a {\it first order} 
differential equation rather than an effective Schr\"odinger 
equation cannot be made.  
For such cases, the exact mode functions themselves are 
used to set up a tunnelling scenario and the resulting 
tunnelling coefficients are interpreted according to 
the tunnelling interpretation to recover the gauge 
invariant result. 
The mixed gauge functions we report in this paper 
have the interesting and useful property that the mode 
functions are combinations of {\it elementary} functions.  
This is in contrast to  the standard modes which are 
transcendental due to the presence of the parabolic 
cylinder functions. 
These new modes are found to be singular on the lightcone. 
This property is very similar to the modes of the 
Schwarzchild-like spacetimes since these too are 
singular on the horizon. 
\par
In classical theory, the action of an uniform electric field 
on a charge imparts a constant uniform acceleration to it.  
The spacetime metric in the rest frame of the charge 
is the Rindler frame.  
Quantum field theory, on the other hand,  predicts particle 
production arising due to the presence of an electric field 
in the spacetime. 
It is therefore of interest to ask if   particle production 
is linked in some way to the the presence of a Rindler 
frame since both are, in a sense, natural for this problem.    
We attempt to link particle production by an uniform 
electric field with processes occuring in the Rindler 
frame by proposing an interpretation of the standard result 
in terms of tunnelling between the two Rindler sectors.
We do this in a heuristic manner and show that this 
tunnelling process between the Rindler sectors gives 
rise to the leading exponential factors in the 
expression for the effective lagrangian (see
Eqn.~(\ref{eqn:efflag}))~\cite{stephens88}. 
\par
The layout of the paper is as follows. 
Section~(\ref{sec:schwarzchild}) contains the 
semi-classical derivation of Hawking radiation without
taking recourse to the Kruskal extension. 
It also contains the reduction to an effective Schr\"odinger 
equation and the subsequent recovery of Hawking radiation 
in the limit of large black hole mass. 
Section~(\ref{sec:electric}) discusses particle production 
in an uniform electric field in a set of new and  
non-trivial gauges. 
It relies heavily on the method of complex paths 
outlined in Appendix~(\ref{sec:facets}). 
In section~(\ref{sec:cpaths}), we attempt to link 
particle production in the electric field to processes 
occuring in the Rindler frame.

\section{Particle Production in spacetimes with  Horizon} \label{sec:schwarzchild}

Hawking's result that a black hole radiates is essentially a 
semi-classical result with the thermal radiation arising because 
of the presence of a horizon in the spacetime structure. 
We will review briefly the conventional derivation of the 
thermal radiation using path integrals.  
Consider a patch of spacetime, which in a suitable co-ordinate 
system, has one of the following forms (we assume $c=1$):
\begin{equation}
ds^2 = B(r)dt^2 - B^{-1}(r) dr^2 - r^2(d\theta^2 + 
\sin^2(\theta)d\phi^2) 
\label{eqn:metric2}
\end{equation}
or
\begin{equation}
ds^2 = B(x)dt^2 - B^{-1}(x) dx^2 - dy^2-dz^2 
\label{eqn:metric3}
\end{equation}
where $B(r)$ and $B(x)$ are functions of $r$ and $x$ respectively.  
The horizon in the above spacetimes is indicated by the surface 
$r=r_0$ ($x=x_0$) at which $B(r)$ ($B(x)$) vanishes.  
We further assume that $B'(r)= dB/dr$ ($B'(x)=dB/dx$) is finite 
and non-zero at the horizon.  
Co-ordinate systems of the form (\ref{eqn:metric2}) can be 
introduced in parts of the Schwarzchild and de-Sitter spacetimes 
while that of the form (\ref{eqn:metric3}) with the choice 
$B(x) = 1 + 2gx$ represents a Rindler frame in flat spacetime.  
Given the co-ordinate system of (\ref{eqn:metric2}) say, in 
some region {\cal R}, we first verify that there is no physical 
singularity at the horizon, which in the case of the 
Schwarzchild black hole, is at the co-ordinate value $r_0=2M$ 
where $M$ is the mass of the black hole.  
Having done that, we extend the geodesics into the past and 
future and arrive at two further regions of the manifold not 
originally covered by the co-ordinate system in (\ref{eqn:metric2}) 
(the Kruskal extension).  
It is now possible to show that the probability for a particle 
with energy $E$ to be lost from the region {\cal R} in relation 
to the probability for a particle with energy $E$ to be gained 
by the region {\cal R} is given by the relation
\begin{equation}
P_{\rm loss} = P_{\rm gain} e^{-\beta E} 
\label{eqn:thermal}
\end{equation}
where $\beta = 8 \pi M$.  This is equivalent to assuming that 
the region {\cal R} is bathed in radiation at temperature 
$\beta^{-1}$.  
In the derivation given in the paper by Hartle and Hawking 
\cite{hawking76}, thermal radiation is derived using the 
semiclassical kernel by an analytic continuation in the 
time co-ordinate $t$ to complex values and it was shown that 
the probability of emission (loss) from the past horizon was 
related to absorption (gain) into the future horizon by the
relation~(\ref{eqn:thermal}).

Since all the physics is contained in the $(t,r)$ plane 
(or the $(t,x)$ plane), we will discuss Hawking radiation 
in $1+1$ dimensions first and show subsequently 
that the  results generalise naturally to 
$3 + 1$ dimensions without modification.
We first derive the semi-classical result in the $(t, r)$ 
(or $(t,x)$) plane by applying a certain prescription to bypass 
the singularity encountered at the horizon.  
After this, we reduce the problem of the Klein-Gordon field 
propagating in the Schwarzchild spacetime to an  effective 
Schr\"odinger problem in (1+1) dimensions and rederive the 
semi-classical result by using the same prescription.

\subsection{Hawking Radiation in 1+1 dimensions}
\label{subsec:hawking}
Consider a certain patch of spacetime in (1+1) dimensions 
which in a suitable co-ordinate system has the line element, 
(with $c=1$) 
\begin{equation}
ds^2 = B(r)dt^2 - B^{-1}(r) dr^2  \label{eqn:metric1}
\end{equation}
where $B(r)$ is an arbitrary function of $r$.   
We assume that the function $B(r)$ vanishes at some $r = r_0$ 
with $B'(r)=dB/dr$ being finite and nonzero at $r_0$. 
The point $r=r_0$ indicates the presence of a horizon.  
It can be easily verified that no physical singularity exists 
at the horizon since the curvature invariants do not have a 
singularity on the horizon. 
Therefore, near the horizon, we expand $B(r)$ as
\begin{equation}
B(r) = B'(r_0)(r - r_0) + {\cal O}[(r-r_0)^2] = R(r_0)(r-r_0).\label{eqn:horizon}
\end{equation}
\par
where it is assumed that $R(r_0) \neq 0$. 
We now use the equation satisfied by the minimally coupled 
scalar field $\Phi$ with mass $m_0$ propagating in 
the spacetime represented by the metric (\ref{eqn:metric1}) to 
obtain the Hamilton-Jacobi equation satisfied by the action 
functional $S_0$.
The semi-classical propagator can be constructed using 
$S_0$ which will be used to analyse the singularity at the horizon. 
(We emphasise again that this method is different from that 
used to compute the Bogolubov coefficients using appropriate 
superpositions of mode functions as outlined in~\cite{bandd82}.)  
\par
The equation satisfied by the scalar field is, 
\begin{equation}
\left(\Box + {m^2_0 \over \hbar^2}\right)\Phi
= 0 \label{eqn:kg}
\end{equation}
where the $\Box$ operator is to be evaluated using metric 
(\ref{eqn:metric1}).  
Expanding the LHS of equation~(\ref{eqn:kg}), 
one obtains, 
\begin{equation}
{1 \over B(r)} {\partial^2 \Phi \over \partial t^2} -
{\partial \over \partial r} \left( B(r) {\partial \Phi 
\over \partial r} \right) = -{m_0^2 \over \hbar^2} \Phi .
\label{eqn:kgr}
\end{equation}
The semiclassical wave functions satisfying the above equation 
are obtained by making the standard ansatz for $\Phi$ which is,
\begin{equation}
\Phi(r,t) = \exp\left[ {i \over \hbar} S(r,t)\right] 
\label{eqn:semi}
\end{equation}
where $S$ is a function which will be expanded in powers of $\hbar$.   Substituting into the wave equation~(\ref{eqn:kgr}), we obtain,
\begin{eqnarray}
\left[ {1 \over B(r)} \left( {\partial S \over \partial t} 
\right)^2 - B(r)\left( {\partial S \over \partial r} 
\right)^2 -m^2_0 \right] && +  \nonumber \\
&& \left({\hbar \over i}\right) \left[ {1 \over B(r)} 
{\partial^2 S \over \partial t^2}- B(r) {\partial^2 S \over 
\partial r^2} -  { dB(r) \over dr}{\partial S \over \partial r} 
\right] = 0 
\label{eqn:seqn1}
\end{eqnarray}
Expanding $S$ in a power series of $(\hbar/i)$,
\begin{equation}
S(r,t) = S_0(r,t) + \left({\hbar \over i}\right) S_1(r,t) +
 \left({\hbar \over i}\right)^2 S_2(r,t) \ldots 
\label{eqn:exp}
\end{equation}
and substituting into Eqn~(\ref{eqn:seqn1}) and neglecting 
terms of order $(\hbar/i)$ and greater, we find to the lowest 
order,
\begin{equation}
{1 \over B(r)} \left( {\partial S_0 \over \partial t} 
\right)^2 - B(r)\left( {\partial S_0 \over \partial r} 
\right)^2 -m_0^2 = 0 
\label{eqn:seqn2}
\end{equation}
Eqn~(\ref{eqn:seqn2}) is just the Hamilton-Jacobi equation 
satisfied by a particle of mass $m_0$ moving in the spacetime 
determined by the metric~(\ref{eqn:metric1}).  
The solution to the above equation is 
\begin{equation}
S_0(r,t) = -Et \pm \int^r {dr \over B(r)} 
\sqrt{E^2 - m^2_0 B(r)} \label{eqn:ssol1}
\end{equation}
where $E$ is a constant and is identified with the energy. 
Notice that in the case of $m_0=0$, Eqn~(\ref{eqn:seqn2}) can 
be exactly solved with the solution
\begin{equation}
S_0 (r,t;m_0=0) = F_1(t - r*) + F_2(t + r*) 
\label{eqn:ssol2}
\end{equation}
where  the ``tortoise'' co-ordinate $r*$ is defined by 
\begin{equation}
r* = \int {dr \over B(r)},
\end{equation}
and $F_1$ and $F_2$ are arbitrary functions.  
If $F_1$ is chosen to be $F_1 = -Et + Er*$ and $F_2$ chosen 
to be $F_2 = -Et - Er*$, then it is clear that the solution 
given in Eqn~(\ref{eqn:ssol2}) is the same as that in 
Eqn~(\ref{eqn:ssol1}) with $m_0$ set to zero.  
Therefore, in the case $m_0=0$, the semiclassical ansatz is exact. 
In the following analysis we will specialise to the case 
$m_0=0$ for simplicity.  The case $m_0\neq 0$ will be 
considered later.  
The essential results do not change in any way.  

The semiclassical 
kernel \, $K(r_2,t_2; r_1,t_1)$ for the particle to propagate 
from $(t_1,r_1)$ to $(t_2,r_2)$ in the saddle point 
approximation can be written down immediately as.
\begin{equation}
K(r_2,t_2; r_1,t_1) = N \exp\left( {i\over \hbar} 
S_0(r_2,t_2; r_1,t_1) \right)
\end{equation} 
where $S_0$ is the action functional satisfying the classical 
Hamilton-Jacobi equation in the 
massless limit and $N$ is a suitable normalization constant.  
$S_0(r_2,t_2; r_1,t_1)$ is given by the relation
\begin{equation}
S_0(r_2,t_2; r_1,t_1)= S_0(2,1) = -E(t_2-t_1) \pm E 
\int^{r_2}_{r_1} {dr \over B(r)}.  
\label{eqn:ssol3}
\end{equation}
The sign ambiguity (of the square root) is related to 
the ``outgoing'' ($\partial S_0 /\partial r\,>0$) or 
``ingoing'' ($\partial S_0 /\partial r\,<0$) nature of the 
particle.
As long as points $1$ and $2$, between which the transition 
amplitude is calculated, are on the same side of the horizon 
(i.e. both are in the region $r>r_0$ or in the region $r<r_0$), 
the integral in the action is well defined and real. 
But if the points are located on opposite sides of the horizon 
then the integral does not exist due to the divergence of 
$B^{-1}(r)$ at $r=r_0$.

Therefore, in order to obtain the probability amplitude for 
crossing the horizon we have to give an extra prescription 
for evaluating the integral \cite{paddy91}.  
Since the point $B=0$ is null we may carry out the calculation 
in Euclidean space or ---equivalently---use an appropriate 
$i\epsilon$ prescription to specify the complex contour over 
which the integral has to be performed around $r=r_0$.  
The prescription we use is that we should take the contour 
for defining the integral to be an infinitesimal semicircle 
{\it above} the pole at $r=r_0$ for outgoing particles on 
the left of the horizon and ingoing particles on the right.  
Similarly, for ingoing particles on the left and outgoing 
particles on the right of the horizon (which corresponds to 
a time reversed situation of the previous cases) the contour 
should be an an infinitismal semicircle {\it below} the 
pole at $r=r_0$. 
Equivalently, this amounts to pushing the singularity at 
$r=r_0$ to $r = r_0 \mp i\epsilon$ where the upper sign 
should be chosen for outgoing particles on the left and 
ingoing particles on the right while the lower sign should 
be chosen for ingoing particles on the left and outgoing 
particles on the right. 
For the Schwarzchild case, this amounts to adding an 
imaginary part to the mass since $r_0 = 2M$.  

Consider therefore, an outgoing particle 
($\partial S_0 /\partial r\,>0$) at $r=r_1<r_0$. 
The modulus square of the amplitude for this particle to 
cross the horizon gives the probability of emission of 
the particle.  
The contribution to $S_0$ in the ranges $(r_1,r_0-\epsilon)$ 
and $(r_0+\epsilon,r_2)$ is real. 
Therefore, choosing the contour to lie in the upper 
complex plane,
\begin{eqnarray}
S_0[{\rm emission}] &=& -E\lim_{\epsilon \to 0} 
\int^{r_0+\epsilon}_{r_0-\epsilon} {dr \over B(r)} 
+ \; ({\rm real \; part})  \nonumber \\
&=& {i \pi E \over R(r_0)}+ ({\rm real \; part})
\end{eqnarray}
where the minus sign in front of the integral corresponds 
to the initial condition that $\partial S_0 /\partial r\,>0$ 
at $r=r_1<r_0$.  
For the sake of definiteness we have assumed $R(r_0)$ in 
Eqn~(\ref{eqn:horizon}) to be positive, so that 
$B(r)<0$ when $r<r_0$.  
(For the case when $R<0$, the answer has to be modified by a 
sign change.)  
The same result is obtained when an ingoing particle 
($\partial S_0 /\partial r\,<0$) is considered at 
$r=r_1<r_0$.  
The contour for this case must be chosen to lie in the 
lower complex plane. 
The amplitude for this particle to  cross the horizon  is 
the same as that of the outgoing particle due to the time 
reversal invariance symmetry obeyed by the system.     

Consider next, an ingoing particle 
($\partial S_0 /\partial r\,<0$) at $r=r_2>r_0$.  
The modulus square of the amplitude for this particle to cross 
the horizon gives the probability of absorption of the 
particle into the horizon. 
Choosing the contour to lie in the upper complex plane, 
we get, 
\begin{eqnarray}
S_0[{\rm absorption}] &=& -E\lim_{\epsilon \to 0} \int^{r_0-\epsilon}_{r_0+\epsilon} {dr \over B(r)} + 
\; ({\rm real \; part})  \nonumber \\
&=& -{i \pi E \over R(r_0)} + ({\rm real \; part})
\end{eqnarray}
The same result is obtained when an outgoing particle 
($\partial S_0 /\partial r\,>0$) is considered at $r=r_2>r_0$.  
The contour for this case should be in the lower complex 
plane and the amplitude for this particle to cross the 
horizon  is the same as that of the ingoing particle due 
to time reversal invariance.

Taking the modulus square to obtain the probability $P$, 
we get,
\begin{equation}
P[{\rm emission}] \propto \exp\left(-{2 \pi E \over \hbar R} 
\right)
\end{equation}
and 
\begin{equation}
P[{\rm absorption}] \propto \exp\left({2 \pi E \over \hbar R} 
\right)
\end{equation}
so that
\begin{equation}
P[{\rm emission}] = \exp\left(-{4 \pi E \over \hbar R} 
\right)P[{\rm absorption}]. \label{eqn:ssys}
\end{equation}
Now time reversal invariance implies that the probability 
for the emission process is the same as that for the absorption 
process proceeding backwards in time and {\it vice versa}. 
Therefore we must interpret the above result as saying that the 
probability of emission of particles is not the same as the 
probability of absorption of particles. 
In other words, if the horizon emits  particles at some time 
with a certain emission probability, the probability of 
absorption of particles at the same time is different from 
the emission probability.  
This result shows that it is more likely for a particular 
region to gain particles than lose them.  
Further, the exponential dependence on the energy allows one to 
give a `thermal' interpretation to this result.  
In a system with a temperature $\beta^{-1}$ the absorption 
and emission probabilities are related by 
\begin{equation}
P[{\rm emission}]= \exp(-\beta E) P[{\rm absorption}] 
\label{eqn:thermalsys}
\end{equation}
Comparing Eqn~(\ref{eqn:thermalsys}) and Eqn~(\ref{eqn:ssys}), 
we identify the temperature of the horizon in terms of $R(r_0)$.
Eqn~(\ref{eqn:ssys}) is based on the assumption that $R>0$.  
If $R<0$ there will be a change of sign in the equation.  
Incorporating both the cases, the general formula for 
the horizon temperature is
\begin{equation}
\beta^{-1} = {\hbar |R|  \over 4\pi} \label{eqn:stemp}
\end{equation}
For the Schwarzchild black hole,
\begin{equation}
B(r) = \left( 1 - {2M \over r}\right) \approx 
{1 \over 2M}(r-2M) + {\cal O}[(r-2M)^2]
\end{equation}
giving $R=(2M)^{-1}$, and the temperature $\beta^{-1}= 
\hbar /8\pi M$.  
For the de-Sitter spacetime,
\begin{equation}
B(r) = (1-H^2r^2) \approx 2H(H^{-1}-r) = -2H(r-H^{-1})
\end{equation}
giving $\beta^{-1}= \hbar H/2\pi$.  
Similarly for the Rindler spacetime
\begin{equation}
B(r) = (1 + 2gr) = 2g(r + (2g)^{-1})
\end{equation}
giving $\beta^{-1}= g\hbar/2\pi$.  
The formula for the temperature can be used for more complicated 
metrics as well and gives the same results as obtained 
by more detailed methods.

The prescription given for handling the singularity is 
analogous to the analytic continuation in time proposed 
by Hawking~\cite{hawking76} to derive Black hole radiance.  
If one started out on the left of the horizon and went 
around the singularity $r=r_0$ by a $2\pi$ rotation instead 
of a rotation by $\pi$, it can be easily shown that it has 
the effect of taking the Kruskal co-ordinates 
$(v,u)$ to $(-v,-u)$.  
A full rotation by $2\pi$ around the singularity can be 
split up into two parts to give the amplitude for emission 
and subsequent absorption of a particle with energy $E$. 
Since the amplitudes for the two processes are not the same 
in the presence of a horizon, one obtains the usual 
Hawking radiation given in Eqn~(\ref{eqn:ssys}) with 
the  value of $R(r_0)$ being $(2M)^{-1}$. 
This process is similar to that given in \cite{hawking76} 
which relates the amplitudes involving the 
past and future horizons.  
In Hawking's paper, analytically continuing the time 
variable $t$ to $t-4Mi\pi$ takes the Kruskal co-ordinates  
$(v,u)$ to $(-v,-u)$ and since the path integral kernel 
is analytic in a strip of $4Mi\pi$ below the real $t$ axis, 
Hawking radiation is obtained by deforming the contour 
of integration appropriately.  
{\it Note however---in our approach---we did not require 
the Kruskal extension and worked entirely in 
the $(t,r)$ co-ordinates}. 

When $m_0 \neq 0$, the validity of the semi-classical 
ansatz must be verified.  
To do this, consider the perturbative expansion~(\ref{eqn:exp}).  
Retaining the terms of order $\hbar/i$ and neglecting 
higher order terms, one finds, upon substituting for $S_0$ 
given by the relation~(\ref{eqn:ssol1}) and solving for $S_1$,
\begin{equation}
S_1 = -E_1t \pm EE_1 \int {dr \over B(r)}
{1\over \sqrt{E^2 - m_0^2 B(r)}} - 
{1 \over 4}\ln(E^2 - m_0^2 B(r)) 
\end{equation}
where $E_1$ is a constant.  From the above equation, it 
is seen that $S_1$ has a singularity of the same 
order as $S_0$ at $r=r_0$.  
When calculating the amplitude  to cross the horizon, 
the contribution from the singular term just appears as 
a phase factor multiplying the semiclassical kernel and 
is inconsequential. 
The non-singular finite terms do contribute to the kernel 
but they contribute the same amount to $S[{\rm emission}]$ 
and $S[{\rm absorption}]$ and they do not affect the relation 
between the probabilities $P[{\rm emission}]$ 
and $P[{\rm absorption}]$. 
Subsequent calculation of the terms $S_2$, $S_3$, and so on, 
show that all these terms have a singularity at the 
horizon of the same order as that of $S_0$. 
Their contribution to the probability amplitude is just 
a set of terms multiplied by powers of $\hbar$ 
which can be neglected.  
From this we can conclude that the semiclassical 
ansatz, in the perturbative limit, is a valid one.
\par
The generalization to (3+1)--dimensions is straightforward. 
We will work with  (\ref{eqn:metric2}) which is in 
spherical polar co-ordinates. 
The results obtained are extendable to (\ref{eqn:metric3}) 
in a straightforward manner.  
The Klein-Gordon equation, written using the 
metric~(\ref{eqn:metric2}), is 
\begin{eqnarray}
{1 \over B(r)} {\partial^2 \Phi \over \partial t^2} - 
{1 \over r^2}{\partial \over \partial r} 
\left( r^2 B(r) {\partial \Phi \over \partial r}\right) 
- && {1 \over r^2 \sin(\theta)}{\partial \over 
\partial \theta}\left( \sin(\theta) {\partial \Phi 
\over \theta} \right) \nonumber \\
&& - {1 \over r^2 \sin^2(\theta)}{\partial^2 
\Phi \over \partial \phi^2}
  = -{m_0^2 \over \hbar^2} \Phi 
\label{eqn:kgr3}
\end{eqnarray}
Since the problem is a spherically symmetric one, one 
can put $\Phi = \Psi(r,t) Y^m_l(\theta, \phi)$ to obtain,
\begin{equation}
{1 \over B(r)} {\partial^2 \Psi \over \partial t^2}  - 
{1 \over r^2}{\partial \over \partial r} 
\left( r^2 B(r) {\partial \Psi \over \partial r}\right) 
+ \left( {l(l+1) \over r^2} + {m_0^2 \over \hbar^2} 
\right)\Psi = 0
\end{equation}
Making the ansatz $\Psi = \exp((i/\hbar) S(r,t))$ 
and substituting into the above equation, we obtain,
\begin{eqnarray}
& &\left[ {1 \over B(r)}\left({\partial S \over 
\partial t}\right)^2 - B(r)\left({\partial S \over 
\partial r}\right)^2 -m_0^2 - {l(l+1) \hbar^2 
\over r^2} \right] \nonumber \\
&& \quad + {\hbar \over i} \left[ {1 \over B(r)}
{\partial^2 S \over \partial t^2} - B(r) 
{\partial^2 S \over \partial r^2} - {1 \over r^2}
{d (r^2 B) \over dr} {\partial S \over \partial r} 
\right] = 0
\end{eqnarray}
Expanding $S$ in a power series as in 
Eqn~(\ref{eqn:exp}), we obtain, to the 
zeroth order in $\hbar/i$, 
\begin{equation}
S_0 = - Et \pm \int^r {dr \over B(r)} 
\sqrt{E^2 - B(r)( m_0^2 + L^2/r^2)} 
\label{eqn:3ssol1}
\end{equation}
where $L^2 = l(l+1)\hbar^2$ is the angular momentum.  
It is easy to see from the above equation that near 
the horizon, the presence of the $L^2$ term can 
be neglected since it is multiplied by $B(r)$. 
Therefore, the  semiclassical result of 
section~(\ref{subsec:hawking}) follows even in the 
case of (3+1)--dimensions. 
The semi-classical ansatz is valid in this case as 
can be seen by calculating explictly the higher 
order terms in the expansion for $S$.  
All these terms have a singularity at the horizon 
of the same order as that of $S_0$ and they contribute 
to the semiclassical propagator either as phase factors 
or as terms multiplied by powers of $\hbar$ 
which are entirely negligible.  
Expanding the Klien-Gordon equation for 
$\Phi$ using the metric~(\ref{eqn:metric3}) 
gives analogous results and will 
not be explictly given.

\subsection{Reduction to an effective Schr\"odinger 
Problem in (1+1) dimensions} 
\label{subsec:schrodinger}
Consider the relativistic equation for the wave 
function $\Phi$ in Eqn~(\ref{eqn:kgr}).  
We include the mass $m_0$ here but  we shall see later that 
it does not appear in the final result. Setting 
\begin{equation}
\Psi(r) \; = \; { e^{-iEt/\hbar} \over \sqrt{B(r)} } Q(r)   
\label{eqn:schsubs}
\end{equation}
we get the equation,
\begin{equation}
 -{d^2 Q(r) \over dr^2} \; - \; \left[ -{B''(r) \over 2 B(r)} 
+ { (B'(r))^2 \over 4 B^2(r)} + { E^2 \over \hbar^2 B^2(r)} - 
{m_0^2 \over \hbar^2 B(r)} \right]\, Q(r) \; = \; 0
\end{equation}
where $B'= (dB/dr)$ and $B'' = (d^2B/dr^2)$.  
Near the horizon, we use the expansion of $B(r)$ given in Eqn~(\ref{eqn:horizon}). Neglecting terms of order 
$1/(r-r_0)$ as compared to terms of order $1/(r-r_0)^2$, 
we get, in the limit of $\hbar \to 0$,
\begin{equation}
 -{d^2 Q(r) \over dr^2} - {g \over (r-r_0)^2} Q(r)  = 0  
\qquad {\rm where} \qquad g\; =\; {E^2 \over \hbar^2 R^2}
\label{eqn:sch2}
\end{equation}
(Notice that $m_0$ does not appear to the leading order 
in the above equation.  
Very close to the horizon, the term containing the mass 
does not contribute significantly.  
Eqn~(\ref{eqn:sch2}) is therefore applicable to both 
massless and massive scalar particles.)   
Making the  transformation $x = (r-r_0)$, we finally 
obtain the effective Schr\"odinger equation for the 
system with $\hbar =2m=1$ with a potential $(-g/x^2)$. 
\begin{equation}
 -{d^2 Q(x) \over dx^2} \; - \; { g  \over x^2} \, Q(x) 
\; = \; 0   
\label{eqn:sch3}
\end{equation} 
This potential is symmetric but singular at the origin $x=0$.  
To make the analogy with the schr\"odinger equation, we will 
replace the righthand side of Eqn.~(\ref{eqn:sch3}) 
by $\widetilde{E} Q(x)$ and finally take the limit of 
$\widetilde{E}\to 0$.  
This ``energy'' $\widetilde{E}$ should not  be confused 
with the energy $E$ of the field in the original 
relativistic system. 
We, therefore,consider  
\begin{equation}
-{d^2 Q(x) \over dx^2} \; - \; { g  \over x^2} \, Q(x)  
\; = \; \widetilde{E}Q(x). 
\label{eqn:sch4}
\end{equation} 
The energy spectrum is continuous for all values of 
$\widetilde{E}$ which, for $\widetilde{E} < 0$, is 
peculiar to this potential since, for energies less 
than the potential energy, the spectrum is usually discrete.  
\par 
The semiclassical analysis follows closely the 
method adopted in section~(\ref{subsec:hawking}).  
The action functional ${\cal A}$ for a classical particle 
moving in a potential $-g/x^2$ satisfies the 
Hamilton-Jacobi equation
\begin{equation}
{\partial {\cal A} \over \partial t}  + \left( { \partial 
{\cal A} \over \partial x} \right)^2  - {g \over x^2} = 0 
\label{eqn:hjeqn}
\end{equation}
The solution can be immediately written down as,
\begin{equation}
{\cal A} = -\widetilde{E} t \pm \int^x \, {dx \over x} 
\sqrt{\widetilde{E} x^2 + g} \label{eqn:semi1}
\end{equation}
Eqn~(\ref{eqn:semi1}) has an integral which is divergent 
if the action is computed for points lying on the 
opposite sides of the horizon $x=0$.  
Since this has a similar form to Eqn~(\ref{eqn:ssol1}), 
the prescription used in evaluating $S[{\rm emission}]$ 
and $S[{\rm absorption}]$ can be similarly used to evaluate 
${\cal A}[{\rm emission}]$ and ${\cal A}[{\rm absorption}]$.  
The results are 
\begin{eqnarray}
{\cal A}[{\rm emission}]&=& i\pi \sqrt{g} + 
({\rm real \; part}) \nonumber \\
{\cal A}[{\rm absorption}]&=& -i\pi \sqrt{g} + 
({\rm real \; part}). 
\end{eqnarray}
Constructing the semiclassical propagator as before 
and taking the modulus square to obtain the probabilities 
for outgoing and ingoing particles, we get
\begin{equation}
P[{\rm outgoing}] = \exp\left[ - {4 \pi   E \over \hbar R} 
\right] P[{\rm ingoing}].
\label{eqn:semisys2}
\end{equation}
The temperature $\beta^{-1}$ for the system is the 
same as that in Eqn~(\ref{eqn:stemp}) and one 
recovers the usual result. 
\par
To verify that the semi-classical analysis is valid, one 
must compute the correction terms and check that these have 
a singularity of the same order as possessed by ${\cal A}$.  
To do this, consider the effective Schr\"odinger 
equation~(\ref{eqn:sch4}) with factors of $\hbar$ put in.
\begin{equation}
-\hbar^2{d^2 Q(x) \over dx^2} \; - \; { g  \over x^2} \, Q(x)  
\; = \; \widetilde{E}Q(x) 
\label{eqn:ssch4}
\end{equation}
Putting $Q(x) = \exp(iA(x)/\hbar)$, and substituting 
into Eqn~(\ref{eqn:ssch4}), \begin{equation}
-i\hbar{d^2 A(x) \over dx^2} + \left({d A(x) \over dx}
\right)^2 = \widetilde{E} + {g \over x^2}. 
\label{eqn:ssch5}
\end{equation}
Expanding $A$ in powers of $\hbar/i$, we get, 
\begin{equation}
A = {\cal A} + {\hbar \over i} A_1 + 
\left({\hbar \over i}\right)^2 A_2 + \ldots 
\label{eqn:ssch6}
\end{equation}
Substituting into Eqn~(\ref{eqn:ssch5}) and proceeding as usual, 
we find that ${\cal A}$ is given by Eqn~(\ref{eqn:semi1}).  
The next term $A_1$ is given by 
\begin{equation}
A_1 = g \int {dx \over x} {1 \over \widetilde{E}x^2 + g}. 
\label{eqn:ssch7}
\end{equation}
The relation for $A_1$ also has a singularity at the origin 
of the same order as ${\cal A}$.  Explicit calculation 
of the subsequent terms in the expansion of $A$ reveals 
that all these terms have a singularity of the same order 
as that of ${\cal A}$ and therefore their net contribution 
to the kernel is either as phase factors or as the exponential 
of finite terms multiplied by powers of $\hbar$. 
Therefore, we conclude as before that the 
semiclassical ansatz is valid.
\par
We show now that the effective Schr\"odinger 
equation in (3+1)--dimensions is the same 
as in Eqn~(\ref{eqn:sch3}).
We consider here the reduction of the Klein-Gordon 
equation in spherical polar co-ordinates obtained 
in Eqn~(\ref{eqn:kgr3}) using 
metric~(\ref{eqn:metric2}).
Setting 
\begin{equation}
\Psi = \exp(-iEt/\hbar) Y^m_l(\theta,\phi) 
\Psi(r) 
\end{equation}
and substituting into Eqn~(\ref{eqn:kgr3}), we obtain,
\begin{equation}
B(r){d^2 \Psi \over dr^2} + {1 \over r^2}{d(r^2B) 
\over dr}{d \Psi \over dr} + \left( {E^2 \over \hbar^2 B(r)} 
- {m^2 \over \hbar^2} - {L^2 \over \hbar^2 r^2} \right) 
\Psi = 0
\end{equation}
Making the substitution
\begin{equation}
\Psi = {1 \over \sqrt{r^2 B(r)}} Q(r)
\end{equation}
we get the result,
\begin{equation}
-{d^2 Q \over dr^2} - \left[ {1 \over B^2}
\left( {(B')^2 \over 4} + {E^2 \over \hbar^2}\right) 
- {1 \over B}\left( {B'' \over 2} +{B'\over r} + 
{m_0^2 \over \hbar^2} + {L^2 \over \hbar^2 r^2}\right) 
\right] Q = 0
\end{equation}
where $B'$ and $B''$ are the first and second 
derivatives of $B(r)$ respectively.  
Near the horizon $r = r_0$, using the expansion 
for $B(r)$ given in Eqn~({\ref{eqn:horizon}), 
it is easy to see that the $1/B^2$ term in the 
above equation dominates over the $1/B$ term which 
can therefore be neglected.  
The resulting Schr\"odinger equation is the same 
as in Eqn~(\ref{eqn:sch3}) in the limit of $\hbar \to 0$.  
It can easily be proved that the effective 
Schr\"odinger equation for the cartesian 
metric~(\ref{eqn:metric3}) gives exactly 
the same result.

\section{Particle production in an uniform Electric field} 
\label{sec:electric}

We now move onto the discussion of particle production in 
an uniform electric field. 
We study a system consisting of a minimally coupled 
scalar field $\Phi$ propagating in flat spacetime in an 
uniform electric field background. 
The two standard gauges, namely the purely time dependent 
and the purely space dependent gauges, are first considered 
and it is shown how the method of complex paths outlined 
in Appendix~(\ref{subsubsec:tunnel}) can be used to construct 
a viable tunnelling interpretation in each case.
The standard quantum field theoretic result will not be 
rederived here since it is well known (though its results 
are used to construct the tunnelling interpretation). 
The tunnelling description is then applied to a few 
non-trivial but simple `mixed' gauges  which are 
functions of both space and time.    
\par
The method of complex paths is a useful tool used to calculate 
the transmission and reflection coefficients in semi-classical 
quantum mechanics in one space dimension~\cite{landau3} 
and we briefly summarise it here, leaving the details 
to Appendix~(\ref{sec:facets}). 
(Readers unfamiliar with this approach should read 
Appendix~(\ref{sec:facets}) at this juncture since 
those results will be used extensively). 
First two disjoint regions are identified where the 
semi-classical wave functions can be written down. 
(The Schr\"odinger potential that will be encountered 
most often will be the inverted harmonic oscillator 
potential $(-x^2)$ which has the semi-classical, disjoint, 
regions located at $x \to \pm \infty$.) 
Then a tunnelling scenario is set up by imposing appropriate 
boundary conditions on the semi-classical wave functions 
in these regions. 
One region is assumed to contain the transmitted wave 
while the other contains the incident and reflected waves. 
To obtain the tunnelling coefficients, the solution is 
analytically continued to the complex plane and the 
behaviour of the  transmitted wave is studied along a 
complex contour (with the space variable now considered 
complex) joining the two regions. 
The contour is chosen such that the semi-classical 
condition is satisfied all along the contour. 
Rotation along the contour transforms the transmitted wave 
either to the incident wave or to the reflected wave 
thus relating the transmission amplitude to either the 
incident or the reflected wave amplitude. 
Using the normalisation condition satisfied by the tunnelling 
coefficients, both coefficients can be determined.  
\par
The complex contour should be chosen so that singularities 
(real and complex) in the potential, where the 
semi-classical ansatz is invalid, are avoided. 
If such singularites exist, they contribute to the 
determination of the tunnelling coefficients. 
Notice that this method works for the exact mode functions 
too (as it must). 
An appropriate tunnelling scenario has to be set up 
and the complex path method can be applied.  
Such situations will be encountered below in certain 
gauges where the explicit modes are known but the problem 
cannot be reduced to an effective Schr\"odinger equation. 

\subsection{Time dependent gauge} 
\label{subsec:electrictime}

The four vector potential $A^{\mu}$ giving rise to a 
constant electric field in the $x$ direction is assumed 
to be of the form
\begin{equation}
A^{\mu} \; = \; (0,\, -E_0t,\, 0,\, 0) \label{eqn:gauge1}
\end{equation}
The electric field is ${\bf E} = E_0\hat{\bf x}$.  
The minimally coupled scalar field $\Phi$ propagating in 
flat spacetime, satisfies the Klein-Gordon equation,
\begin{equation}
\left( (\partial_{\mu} \; + \; iqA_{\mu})(\partial^{\mu} 
\; + \; iqA^{\mu}) + m^2\right)\Phi \; = \; 0 
\label{eqn:electric11}
\end{equation}
where $m$ is the mass and $q$ is the charge of the field. 
The mode functions of $\Phi$ can be expressed in the form 
$\Phi(t,{\bf x}) = f_{\bf k}(t)e^{i{\bf k}\cdot {\bf x}}$ 
where $f_{\bf k}(t)$ satisfies the equation,
\begin{equation}
{d^2 \over dt^2}f_{\bf k} \; + \; \left[ m^2 + k_{\perp}^2 + 
(k_x + qE_0t)^2\right]f_{\bf k} \; = \; 0 \; ;\qquad {\bf k}_{\perp} 
\; = \; (k_y,k_z) .
\label{eqn:electric12}
\end{equation}
Introducing the variables,
\begin{equation}
\tau \; = \; \sqrt{qE_0}\, t \; + \; {\omega \over \sqrt{qE_0}}
\qquad ;\qquad  \lambda \; =\; {k_{\perp}^2 + m^2 \over  qE_0} 
\label{eqn:electric13}
\end{equation}
we obtain the equation,
\begin{equation}
-{d^2 \over d\tau^2}f_{\bf k} \; - \; \tau^2 f_{\bf k} 
\; = \; \lambda f_{\bf k} 
\label{eqn:electric14} 
\end{equation}
The above equation is essentially a Schr\"odinger equation in an 
inverted oscillator potential with a positive ``energy'' $\lambda$. 
Since the energy is positive, the problem is essentially an 
{\it over the barrier reflection} problem.  Using the results of section~(\ref{subsubsec:potential1}), we can calculate the 
reflection and transmission  coefficients exactly as
\begin{equation}
R \; = \;  {e^{-\pi\lambda} \over 1 + e^{-\pi\lambda} } \qquad ; 
\qquad T \; = \; {1 \over 1 + e^{- \pi\lambda} } 
\label{eqn:electric15}
\end{equation}
where we have put $\hbar = 2m = g_1= 1$ and set 
$E_1 = \lambda$ in Eqns.~(\ref{eqn:electpot5}, 
\ref{eqn:electpot6}). 
To identify the Bogoliubov coefficients $\alpha_{\lambda}$ 
and $\beta_{\lambda}$, we recast the normalisation condition 
$R+T=1$ in the form,
\begin{equation}
{1 \over T} \; - \; {R \over T} \; = \; 1
\end{equation}
and then identify $|\beta_\lambda|^2$ with $R/T$ and 
$|\alpha_\lambda|^2$ with $1/T$. 
Therefore, the Bogoliubov coefficients are given by, 
\begin{eqnarray}
|\beta_\lambda|^2 \; &=& \; e^{-\pi\lambda} \; = \; 
\exp\left(-{\pi(k_{\perp}^2 + m^2) \over qE_0}\right) \nonumber \\
|\alpha_\lambda|^2 \; &=& \;   e^{-\pi\lambda} \; + \; 1 \; = \; 
\exp\left(-{\pi(k_{\perp}^2 + m^2)\over qE_0}\right) \; + \; 1 
.\label{eqn:electric16}
\end{eqnarray}
The transmission and reflection coefficients are time 
reversal invariant and are dependent only on the energy 
(magnitude and sign). 
They are also independent of the direction 
in which the boundary conditions are applied. 
To obtain a dynamical picture of particle production, 
we have to interpret these quantities suitably. 
In the present case, the following interpretation seems adequate. 
A purely positive frequency wave with amplitude square $T$ 
in the infinite past, $t \to -\infty$, evolves into a 
combination of positive and negative frequency waves in 
the infinite future $t \to \infty$ with the negative frequency 
waves having an amplitude square $R$ and the positive frequency 
waves having an amplitude unity. 
The quantity  $R/T$  determines the overlap between the 
negative frequency modes in the distant future and the 
positive frequency modes in the distant past (the notation 
here differs from the treatment given in \cite{paddy91}, 
\cite{brout95}). 
This is identified with the modulus square of the 
Bogoliubov coefficient $\beta_{\lambda}$ which is the 
particle production per mode $\lambda$. 
Using the normalisation condition satisfied by the Bogoliubov 
coefficients, $|\alpha_\lambda|^2 - |\beta_{\lambda}|^2 = 1$,
$|\alpha_\lambda|^2$ can be calculated to be $1/T$.
Once the Bogoliubov coefficients have been identified, the 
effective lagrangian can be easily calculated.  
This derivation will not be repeated here.  
We refer the reader to Ref.~\cite{brout95} and 
Ref.~\cite{paddy91} for the explicit calculation.
Note that the particular interpretation given in this 
case is due to its similarity with the more rigourous 
calculation by quantum field theoretic methods. 
In the next section, in which we discuss the space dependent 
gauge, we will be forced to adopt a different interpretation 
in order to identify particle production.

\subsection{Space dependent gauge} 
\label{subsec:electricspace}
The four vector potential $A^{\mu}$ giving rise to a constant 
electric field in the $x$ direction is now assumed to be 
of the form
\begin{equation}
A^{\mu} \; = \; (-E_0x, \, 0,\, 0,\, 0) \label{eqn:gauge2}
\end{equation}
The electric field is ${\bf E} = E_0\hat{\bf x}$ as before.  
The field  $\Phi$ satisfies Eqn.~(\ref{eqn:electric11}) as 
before.
Substituting for the potential $A^{\mu}$ from 
Eqn.~(\ref{eqn:gauge2}) into Eqn.~(\ref{eqn:electric11}), 
we obtain,
\begin{equation}
\left( \partial_t^2 \; - \; \nabla^2 \; - \; 2iqE_0x\partial_t 
\; - \; q^2E_0^2x^2 \; + \; m^2\right)\Phi \; = \; 0 
\label{eqn:electric22}
\end{equation}
We write $\Phi$ in the form
\begin{equation}
\Phi \; = \; e^{-i\omega t} \, e^{i k_y y + ik_zz} \, \phi(x) 
\label{eqn:electric23}
\end{equation}
and obtain the differential equation satisfied 
by $\phi$ as
\begin{equation}
{d^2 \phi \over dx^2} \; + \; \left( (\omega + qE_0x)^2 -  
k_{\perp}^2 - m^2 \right)\phi \; = \; 0
\label{eqn:electric24}
\end{equation}
where we have used the notation $k_{\perp}^2 = k_y^2 + k_z^2$. 
Making  the following change of variables 
\begin{equation}
\rho  \; = \; \sqrt{qE_0}x \; + \; {\omega \over \sqrt{qE_0}} 
\qquad ; \qquad
\lambda \; = \; {k_{\perp}^2 + m^2 \over qE_0}
\label{eqn:electric25}
\end{equation}
into the differential equation for $\phi$, it reduces 
to the form,
\begin{equation}
-{d^2 \phi \over d\rho^2} \; - \; \rho^2\phi \; = \; 
-\lambda\phi 
\label{eqn:electric26}
\end{equation}
In this form, we see that the above differential equation 
has the form of an effective Schr\"odinger equation with 
an inverted harmonic oscillator potential and an 
negative energy $-\lambda$.  
If we apply the results of section~(\ref{subsubsec:potential1}), 
we obtain the result for {\it tunnelling through the barrier}.  
Following the treatment in Ref.~\cite{paddy91} and using 
the results of section~(\ref{subsubsec:potential1}) we can 
calculate the reflection and transmission  coefficients exactly 
as
\begin{equation}
R \; = \;  {e^{ \pi\lambda} \over 1 + e^{\pi\lambda} }\qquad ; 
\qquad T \; = \; {1 \over 1 + e^{\pi\lambda} } 
\label{eqn:electric27}
\end{equation}
where we have put $\hbar = 2m = g_1 = 1$ and set 
$E_1 = -\lambda$ in Eqns.~(\ref{eqn:electpot5}, 
\ref{eqn:electpot6}). 
We cast the renormalisation condition $R+T=1$ in the form
\begin{equation}
{1 \over R} \; - \; {T\over R} \; = \; 1
\end{equation}
and then identify the rate of particle production per mode with 
$T/R$. 
The interpretation of particle production using the tunnelling 
picture now proceeds as follows. 
A right moving travelling wave of amplitude square $1/R$ is 
incident on the potential. 
A fraction $T/R$ is transmitted through it and a wave of unit 
amplitude is scattered back.  
The tunnelling probability, which is $T/R$, is interpreted as 
the rate at which particles are being produced by the 
background electric field. 
This matches exactly with the expression for 
$|\beta_{\lambda}|^2$ given in Eqn.~(\ref{eqn:electric16}). 
With this interpretation, we recover the usual gauge 
independent result.  

\subsection{Mixed gauges} 
\label{subsec:electricmixed}
We shall now study the problem in a new set of gauges 
which prove to be useful and instructive. 
(As far as the authors know these gauges have not been 
studied in the literature before.)
When parameters which specify these gauges are varied, 
the problem can be mapped either onto a  ``tunnelling 
through the barrier'' or ``over the barrier 
reflection'' system. 
Then, using the tunnelling description developed previously 
for each of these systems, the gauge invariant result 
is obtained.  
\par
For certain ranges of these parameters, the scalar wave 
equation can be reduced to solving a  second order 
equation which can be converted to an effective 
Schr\"odinger equation.  
This equation is studied using the complex path method 
as in section~(\ref{subsec:electrictime}) and 
section~(\ref{subsec:electricspace}).  
The solutions to these effective Schr\"odinger equations 
are usually transcendental in form.  
\par
For other parameter ranges, however, the scalar field equation 
reduces to a {\it first} order differential equation 
whose solution is a combination of elementary functions.  
To recover the gauge invariant result, the solution and 
its complex conjugate are used to set up a tunnelling 
scenario from which the tunnelling coefficients are obtained. 
Then, using the tunnelling interpretation, the gauge 
invariant result of particle production is recovered.

\subsubsection{Gauge Type 1}
\label{subsec:electrictype1}
The first gauge type we consider is a simple generalisation 
of the space-dependent gauge (in Eqn.~(\ref{eqn:gauge2})) 
of the form
\begin{equation}
A^{\mu} = (-E_0x + D_1t, \, 0,\, 0,\, 0) \label{eqn:gauge3}
\end{equation}
where $E_0$ is as usual the magnitude of the electric field 
and $D_1$ is a gauge parameter which can be positive or negative.  
It is easily verified that the  above gauge type gives the 
same electric field as the pure time and space gauges did. 
The differential equation for the scalar field  $\Phi$ is,
\begin{equation}
\left( \partial_t^2 \; - \; \nabla^2 \; + \; 
2iq(-E_0x + D_1t)\partial_t \; - \; q^2(-E_0x+D_1t)^2 + iqD_1 
+  m^2\right)\Phi \; = \; 0
\label{eqn:electric31}
\end{equation}
We make a judicious choice of variables of the form
\begin{equation}
u \; = \; -E_0x \; + \; D_1t \qquad ; 
\qquad v \; = \; E_0x \; + \; D_1t .
\label{eqn:electric32}
\end{equation}
Notice that the coefficients in Eqn.~(\ref{eqn:electric31}) 
are dependent only on the variable $u$ and not on 
$v$, $y$ or $z$ and so the above equation is seperable 
in the variables $(u,v,y,z)$. 
Expressing the derivatives $(\partial_t, \partial_x)$ in 
terms of $(\partial_u, \partial_v)$ and 
writing $\Phi$ in the form
\begin{equation}
\Phi \; = \;  e^{i k_y y + ik_zz} e^{-i\gamma v} 
\phi(u),
\label{eqn:electric32a}
\end{equation}
one obtains 
\begin{eqnarray}
(D_1^2 - E_0^2){d^2 \phi \over d u^2} \; &+& \; 
2i(qD_1u - \gamma (D_1^2 + E_0^2)){d\phi \over du} 
\nonumber \\
&& \quad + \; (m^2 + k_{\perp}^2 + iqD_1 -\gamma^2 
(D_1^2 - E_0^2) - q^2u^2 + 2qD_1\gamma u )\, \phi(u) 
\; = \; 0
\label{eqn:electric33}
\end{eqnarray}
It is easy to see from the above equation that two distinct 
cases can be identified here, namely, $|D_1| \neq |E_0|$ 
and $D_1 = \pm E_0$. 
In the first case, the differential equation is a second 
order one and the effective Schrodinger equation can be 
obtained by eliminating the first derivative. 
In the second case, however, the resulting equation is 
a {\it first order} differential equation whose 
solution is an elementary function.    
\par
Consider first the case $|D_1| \neq |E_0|$. 
Writing $\phi(u)$ in the form,
\begin{equation}
\phi(u) \; = \; Q(u) \, \exp-i\left[{qD_1 
\over 2(D_1^2 - E_0^2)}\, u^2 \,-\,
{\gamma (D_1^2 + E_0^2) \over (D_1^2 - E_0^2)}\, u 
\right]
\label{eqn:electric34}
\end{equation}
and defining a new variable $\rho = u - (2\gamma D_1/q)$, 
one obtains the effective Schr\"odinger equation as,
\begin{equation}
-{d^2 Q \over d\rho^2}\; - \; {q^2E_0^2 \rho^2 \over 
(D_1^2 - E_0^2)^2}\, Q(\rho) \; = \; 
{(m^2 + k_{\perp}^2) \over (D_1^2 - E_0^2)}\, Q(\rho) .
\label{eqn:electric36}
\end{equation}
The effective potential is clearly seen to be that of 
an inverted oscillator.   
Notice that, depending on whether $D_1^2 < E_0^2$  or 
$D_1^2 > E_0^2$, the problem reduces to a 
``tunnelling through the barrier'' problem or an 
``over the barrier reflection'' problem respectively. 
Using the results of section~(\ref{subsubsec:potential1}) 
and especially Eqn.~(\ref{eqn:electpot3}), the value of 
the quantity $2\varepsilon_1$ is found to be 
(with $\hbar = 2m = 1$)
\begin{equation}
2\varepsilon_1 \; = \; \sqrt{1 \over g_1} E_1 \; 
= \; {k_{\perp}^2 + m^2 \over  qE_0} Sgn(D_1^2 - E_0^2) \; 
= \; \lambda \, Sgn(D_1^2 - E_0^2)
\label{eqn:electric37}
\end{equation}
where $Sgn(x)$ is the sign function which is positive 
for $x>0$ and negative for $x<0$ and $\lambda$ is 
defined in Eqn.~(\ref{eqn:electric14}). 
Thus, the tunnelling coefficients are given by 
either Eqn.~(\ref{eqn:electric15}) or 
Eqn.~(\ref{eqn:electric27}) depending on 
$Sgn(D_1^2 - E_0^2)$ but are neverthless independent of 
$D_1$ as expected. 
The tunnelling interpretation required to recover the 
standard result proceeds accordingly.  
Note that only the $Q(u)$ part of the full solution to 
$\Phi$ contributes to the transmission and reflection 
coefficients. 
The solutions to $Q(u)$ are the usual parabolic 
cylinder functions. 
\par
Now, consider the more interesting case $D_1 = +E_0$ 
with $u=E_0(t-x)$. 
The differential equation Eqn.~(\ref{eqn:electric33}) 
for $\phi$ reduces to
\begin{equation}
2i(qE_0u - 2\gamma  E_0^2){d\phi \over du} \; 
+ \; (m^2 + k_{\perp}^2 + iqE_0  - q^2u^2 + 2qE_0\gamma u )\, 
\phi(u) \; = \; 0
\label{eqn:electric38}
\end{equation}
The solution is easily obtained to be
\begin{equation}
\phi(u) \; =  \left[ \sqrt{q \over E_0}\, u - 
2\gamma \sqrt{E_0 \over q}  \right]^{i\lambda/2 \,-\, 1/2} 
\, e^{-iqu^2/4E_0}
\label{eqn:electric39}
\end{equation}
with $\lambda$ being defined in Eqn.~(\ref{eqn:electric25}). 
This solution resembles the asymptotic forms for some of 
the parabolic cylinder functions except that it is 
{\it exact} and is clearly a combination of elementary functions. 
(Notice that the solution is singular on the  surface
$t - x  = 2\gamma /q$ which is remniscent of the behaviour 
of black hole modes near the horizon.  
The implications of this will be discussed in a future publication.)
To recover the gauge invariant result in this case, notice 
that the complex conjugate  $\phi^*(u)$ is also a solution.  
With this pair of independent modes, one can apply the 
theory given in section~(\ref{subsubsec:tunnel}) to set up 
a tunnelling problem with the appropriate boundary 
conditions at $u = \pm \infty$.  
This is most conveniently done by defining a new 
dimensionless variable 
\begin{equation}
s \; = \; \sqrt{q \over E_0}\, u 
\label{eqn:electric310}
\end{equation}
The mode function $\phi$ now becomes
\begin{equation}
\phi (s) \; = \;  \left(s - 2\gamma 
\sqrt{E_0/q}\right)^{i\lambda/2 \,-\, 1/2} \, 
e^{-is^2/4}
\label{eqn:electric311} 
\end{equation}
with an analogous expression for $\phi^*(s)$. 
We now assume that the wave function is a 
right moving travelling wave $\phi_R$ as 
$s \to + \infty$ while as $s\to -\infty$, 
it is a superposition of an incident wave 
of unit amplitude and a reflected wave given by $\phi_L$.  
Therefore, we have, 
\begin{eqnarray}
\phi_R = C_3 s^{-i\lambda/2 - 1/2} e^{is^2/4} 
\qquad \qquad \qquad \qquad \qquad \; && 
\qquad \qquad (s \to +\infty) \nonumber  \\
\phi_L =  (-s)^{i\lambda/2 - 
1/2} e^{- is^2/4} + 
C_2 (-s)^{-i\lambda/2 - 1/2} e^{is^2/4} && 
\qquad \qquad (s \to -\infty) 
\label{eqn:electric312}
\end{eqnarray}
where $C_3$ and $C_2$ are the transmission and reflection 
amplitudes respectively. 
We have made the approximation 
$s - 2\gamma \sqrt{E_0/q}\approx s$ which holds in 
the limit $s\to \pm \infty$.  
Applying the method of complex paths (see 
Appendix~(\ref{sec:facets})) to rotate $\phi_R$ about 
the {\it upper} complex contour (in accordance with the 
semi-classical ansatz), one obtains
\begin{equation}
C_2 \; = \; -i C_3 \, \exp\left({\pi \lambda \over 2}
\right) 
\label{eqn:electric313}
\end{equation}
Using the normalisation condition $|C_2|^2 + |C_3|^2 =1$, 
we get,
\begin{equation}
R \; = \; |C_2|^2 \; = \; {e^{\pi\lambda} \over 1 + 
e^{\pi\lambda} } \qquad ; 
\qquad T \; = \; |C_3|^2 \; = \; {1 \over 1 + 
e^{\pi\lambda} }
\label{eqn:electric314}
\end{equation}
which are the usual tunnelling coefficients given in 
Eqn.(\ref{eqn:electric27}). 
It is therefore seen that the system corresponds to an 
tunnelling through the barrier problem. 
The interpretation, of course, follows that given 
in section~(\ref{subsec:electricspace}).  
\par
For the case $D_1 = -E_0$, the modes are given by
\begin{equation}
\phi(u) \; =  \left[ \sqrt{q \over E_0}\, u + 
2\gamma \sqrt{E_0 \over q}  \right]^{-i\lambda/2 \,-\, 1/2} 
\, e^{iqu^2/4E_0}
\label{eqn:electric315}
\end{equation}
An analysis similar to the case $D_1 = +E_0$ shows that the 
tunnelling coefficients are the same as 
in  Eqn.~(\ref{eqn:electric314}).  
This system also corresponds to tunnelling through the 
barrier with the corresponding interpretation given 
in section~(\ref{subsec:electricspace}) required in order 
to recover paticle production. 
Therefore, it is seen that, for the case $|D_1| \neq |E_0|$, 
the magnitude of $D_1$ decides the appropriate tunnelling 
interpretation even though it does not appear in the 
final expressions for the tunnelling coefficients. 
In contrast, for the cases $D_1 = \pm E_0$, the system 
reduces to a tunnelling through the barrier with the 
sign of $D_1$ not playing any role in deciding the 
appropriate interpretation.   
\par
The above considerations also work for an analogous 
generalisation of the time-dependent gauge of the form
\begin{equation}
A^{\mu} = (0, -E_0t + D_1x, \, 0,\, 0,\, 0) 
\end{equation}
in an obvious manner and we will not repeat the discussion.  

\subsubsection{Gauge Type 2}
\label{subsec:electrictype2}
The second gauge type we consider is of the form,
\b
A^{\mu} = ( D_1 t - D_2x,\, D_2x - D_1t,\, 0,\, 0) 
\label{eqn:gauge4} 
\e
where $D_1$ and $D_2$ are arbitrary constants such that 
$E_0 = D_1 + D_2$. 
The  magnitude and direction of the uniform electric field 
is seen to be the same. 
It is in this gauge that, for $D_1 = D_2$, the mode functions 
have the simplest form possible. 
Writing out the differential equation for the scalar 
field $\Phi$ and defining new variables
\begin{equation}
u = D_1t - D_2x \qquad ; \qquad v = D_1t + D_2x 
\label{eqn:electric41}
\end{equation}
and setting $\Phi = e^{i k_y y + ik_zz}\, e^{-i\gamma v}\, 
\phi(u)$
as before, one obtains,
\begin{eqnarray}
(D_1^2 - D_2^2)\, {d^2\phi \over du^2} \; 
&+& \; 2i\left[q(D_1 + D_2)u - \gamma (D_1^2 + D_2^2)\right]\, 
{d\phi \over du} \nonumber \\
 &+& \; \left[ m^2 + k_{\perp}^2 + iq(D_1+D_2) - 
\gamma^2 (D_1^2 - D_2^2) + 2\gamma q(D_1-D_2)u\right] 
\phi (u) \; = \; 0
\label{eqn:electric42}
\end{eqnarray}
Here too, two distinct cases can be distinguished, namely, 
$D_1 \neq D_2$ and $D_1 = D_2$ (we will discuss the case 
$D_2 = -D_1$, which corresponds to a `pure gauge' with 
zero electric field a little later).
\par
Consider first the case $D_1 \neq D_2$. 
Writing $\phi$ in the form,  
\begin{equation}
\phi (u) \; = \; \exp-i\left( {q \over 2(D_1 - D_2)} u^2 - 
{\gamma (D_1^2 + D_2^2) \over (D_1^2 - D_2^2)} u \right) Q(u) .
\label{eqn:electric43}
\end{equation}
and introducing a new variable 
$\rho = u - (2\gamma D_1D_2/qE_0)$, one obtains an 
effective Schr\"odinger equation as,  
\begin{equation} 
-{d^2 Q \over d\rho^2}\; - \; {q^2E_0^2 \rho^2 \over 
(D_1^2 - D_2^2)^2}\, Q(\rho) \; = \; 
{(m^2 + k_{\perp}^2) \over (D_1^2 - D_2^2)}\, Q(\rho) . 
\label{eqn:electric44}
\end{equation}
where we have used $E_0 = D_1 + D_2$.  
This equation has the same form as that in  
Eqn.~(\ref{eqn:electric36}). 
Depending on whether $D_1^2 < D_2^2$  or $D_1^2 > D_2^2$, 
the problem reduces to a ``tunnelling through the barrier'' 
problem or an ``over the barrier reflection'' problem 
respectively. 
Using the results of section~(\ref{subsubsec:potential1}) 
and especially Eqn.~(\ref{eqn:electpot3}), one finds 
the value of the quantity $2\varepsilon_1$ to be 
(with $\hbar = 2m = 1$)
\begin{equation}
2\varepsilon_1 \; = \; \sqrt{1 \over g_1} E_1 \; = 
\; {k_{\perp}^2 + m^2 \over  qE_0} Sgn(D_1^2 - D_2^2) 
\; = \; \lambda \, Sgn(D_1^2 - D_2^2)
\label{eqn:electric45}
\end{equation}
Thus, the tunnelling coefficients are given by 
either Eqn.~(\ref{eqn:electric15}) or 
Eqn.~(\ref{eqn:electric27}) depending on 
$Sgn(D_1^2 - D_2^2)$ and are dependent 
only on  $E_0$ as expected. 
The tunnelling interpretation required to recover the 
standard result proceeds accordingly. 
\par
However, when $D_1 = D_2$, the differential equation 
for $\phi(u)$ reduces to
\begin{equation}
4i(qD_1u - \gamma D_1^2){d\phi \over du} \; + \; 
(m^2 + k_{\perp}^2 + 2iqD_1) \phi (u) \; = \; 0
\label{eqn:electric46}
\end{equation}
with the solution
\begin{equation}
\phi (u) = \left( \sqrt{2q\over E_0}\, u - 
\gamma \sqrt{E_0\over 2q}\, \right)^{i\lambda/2 \, - \, 1/2} .
\label{eqn:electric47}
\end{equation}
This is the simplest mode function possible for 
any gauge of the electric field.  
This solution is also singular just like the modes 
in Eqn.~(\ref{eqn:electric39}) but on the surface 
$t-x = \gamma/q$. 
The form of the function $\phi (u)$ clearly indicates the 
factor responsible for the non-zero tunnelling coefficients.  
One can, in a manner similar to that  outlined for the modes~(\ref{eqn:electric39}), use the independent mode 
functions $\phi$ and $\phi^*$ to set up a tunnelling 
problem in the limit $u\to \pm \infty$. 
Rotating the transmitted wave in the {\it upper} complex 
plane (this is the only contour possible in accordance 
with the semi-classical ansatz) the tunnelling coefficients 
are found to be those in Eqn.~(\ref{eqn:electric15}). 
The system, in this case, corresponds to over the barrier 
reflection and the interpretation required to obtain 
particle production follows that in 
section~(\ref{subsec:electrictime}). 
\par
Finally, we look at the case $D_2 = -D_1$. The electric field, 
in this pure gauge, is identically zero. 
Solving Eqn.~(\ref{eqn:electric42}) for this case, the solution 
for $\phi$ is found to be 
\begin{equation}
\phi (u) \; = \;  e^{-i\lambda u/2\gamma D_1}\, 
e^{-iqu^2/2D_1} 
\end{equation}
It is clear from the above form of $\phi$ that the 
transmission coefficient is unity and the reflection 
coefficient is zero. 
No particle production takes place. 
Explicit calculations, of course, verify this. 
The tunnelling interpretation gives a null result in 
this case.
\par
The mixed gauge types in Eqn.~(\ref{eqn:gauge3}) and 
Eqn.~(\ref{eqn:gauge4}) represent the same electric field. 
Hence, these gauges must be related to the pure space and 
time gauges in Eqns~(\ref{eqn:gauge1}, \ref{eqn:gauge2}) 
by gauge transformations of the form 
$\tilde{A}^{\alpha} = A^{\alpha} + \partial^{\alpha} f$ 
where $f$ is a suitable gauge function. 
Choosing the base gauge $A^{\alpha}$ as the space-dependent 
gauge in Eqn.~(\ref{eqn:gauge2}), one obtains the 
following gauge functions for the two mixed gauge types. 
\begin{eqnarray}
A^{\alpha} \; \to \; \tilde{A}^{\alpha} \; &=& 
\; (-E_0x + D_1t, \, 0,\, 0,\, 0) \qquad \; \qquad 
\qquad f \; = \; {1 \over 2}\, D_1 t^2 
\nonumber \\
A^{\alpha} \; \to \; \tilde{A}^{\alpha} \; &=& 
\; ( D_1 t - D_2x,\, D_2x - D_1t,\, 0,\, 0) \;\;\, 
\qquad f  \; = \; D_1tx \; + \; {1\over 2} \, 
\left( D_1t^2\; - \; D_2x^2 \right) 
\end{eqnarray}
The gauge transformed scalar fields in the 
$\tilde{A}^{\alpha}$ gauge ought to be related to 
those in the $A^{\alpha}$ gauge by a phase factor $e^{if}$. 
But the solutions give in Eqn.~(\ref{eqn:electric36}) 
(with $D_1 = E_0$) and   Eqn.~(\ref{eqn:electric47}) 
(with $D_1 = D_2 = E_0/2$) are clearly  not gauge 
transformations of the mode functions of 
gauge~(\ref{eqn:gauge2}) which contain parabolic 
cylinder functions. 
These modes are intrinsic to the gauges themselves 
and cannot be obtained by a simple gauge transformation of 
the modes of the space-dependent gauge. 
This is reminiscent of the situation in the Rindler 
and Minkowski frames.  
The Rindler plane wave modes of the form 
$e^{-i\omega \tau}f(\xi)$ (where $\tau$ and $\xi$ are 
the time and space co-ordinates in the Rindler frame) are 
{\it not} obtained from the Minkowski plane wave modes of 
the form $e^{-i\omega t + kx}$ (where $t$ and $x$ are 
the Minkowski time and space co-ordinates) by a 
co-ordinate transformation.   
\par
The solutions given in Eqn.~(\ref{eqn:electric36}) 
and Eqn.~(\ref{eqn:electric47}) can be obtained in a 
more general and elegant fashion as follows.  
Consider the Minkowski metric expressed in the $(u,v,y,z)$ 
co-ordinate system where $u=t-x$ and $v=t+x$.
\begin{equation}
ds^2 \; = \; dudv \; - \; dy^2 \; - \; dz^2
\label{eqn:electric60}
\end{equation}
Assume a vector potential of the form
\begin{equation}
A^{\mu} \; = \; (g(u),\, h(u), \, 0,\, 0)
\label{eqn:electric61}
\end{equation}
where $g(u)$ and $h(u)$ are arbitrary functions with 
$A^{\mu} = (A^u,A^v,A^y,A^z)$ being expressed in 
the $(u,v,y,z)$ co-ordinate system  and {\it not} 
in the $(t,x,y,z)$ system. 
The electric field is given by 
${\bf E} = (\partial g(u)/\partial u) \hat{\bf x}$.  
Notice that the $v$ component of the vector potential, 
$A^v= h(u)$,  does not contribute to the electric field 
and is a `pure' gauge function. 
Writing down the scalar wave equation in the $(u,v,y,z)$ 
system and setting 
$\Phi =  e^{i k_y y + ik_zz} e^{-i\gamma v} \phi(u)$, 
one obtains the general solution for $\phi (u)$ as
\begin{equation}
\ln(\phi(u)) \; = \; -{k_{\perp}^2 +m^2 \over 2i} 
\int \! {du \over qg(u) - 2\gamma} \; - \; {1\over 2}
\ln\left(qg(u) - 2\gamma\right) \; + \; {i\over 2} 
\int \! du h(u)
\label{eqn:electric62})
\end{equation}
For the uniform electric field, $g(u) = E_0 u + C_0$ 
with $C_0$ being a constant. 
Both the gauges in Eqn.~(\ref{eqn:gauge3}) 
(with $D_1 = E_0$) and Eqn.~(\ref{eqn:gauge4}) 
(with $D_1 = D_2 = E_0/2$) have a similar form for $g(u)$, 
but differ in the form of $h(u)$. 
To set up a tunnelling scenario the explicit form of 
$h(u)$ must be known as this will determine whether 
the system is a tunnelling through the barrier or an 
over the barrier reflection system.  
For more general forms for $g(u)$, it is clear that the 
gauge invariant quantity $E^2 - B^2 \neq 0$ which implies 
non-zero particle production.  
This is related to the pole structure of the first term 
in Eqn.(\ref{eqn:electric62}) and will be explored 
further in a future publication.
\par
In summary, it is seen that by a judicious choice of 
interpretation of the transmission and reflection coefficients 
in each of the two pure gauges, the standard gauge invariant 
result can be obtained. 
This tunnelling interpretation works in the case of the 
mixed gauges too with the gauge parameters deciding whether 
the system is an `over the barrier reflection' one or 
a `tunnelling through the barrier' one. 
The effective Schr\"odinger equation that is analysed in 
these gauges is expressed in suitable variables that are 
combinations of the usual spacetime variables. 
In cases where there is no explicit reduction to an 
effective Schr\"odinger form, the exact modes themselves 
are used to set up a tunnelling scenario with the recovery 
of the tunnelling coefficients. 
The two mixed gauges that were considered lend themselves 
to easy solutions but more complicated ones may be 
considered in a similar manner. 
The mode functions in Eqn.~(\ref{eqn:electric39}) 
and Eqn.~(\ref{eqn:electric47}) are a combination of 
elementary functions unlike the previously known modes. 
The tunnelling coefficients arise solely due to the 
presence of a factor $s^{\pm i\lambda/2}$ in the 
semi-classical wave functions and in the 
exact mode functions.  
It is interesting to note that, in the case of black hole 
spacetimes too, it is precisely a factor of the form 
$s^{\pm i\varepsilon}$ near the horizon that gives 
rise to the non-trivial result of Hawking radiation.   
\par
In the last section, we go on to interpret particle 
production by an electric field in terms of tunnelling 
between the two sectors of the Rindler spacetime.   

\section{Complex paths interpretation of particle 
production in Electric field} 
\label{sec:cpaths}
In section~(\ref{sec:electric}), we noted how particle 
production can be calculated using 
the tunnelling interpretation.  
This interpretation gives the same result in both the 
space as well as the time dependent gauges. 
The spectrum of particles produced by the electric field 
is not thermal in contrast to the spectrum 
seen by a Rindler observer. 
We use the formal tunnelling method to 
show, in a heuristic manner, how this particle 
production can be obtained by tunnelling between the 
two sectors of the Rindler spacetime.  

The imaginary part of the effective lagrangian 
${\rm Im L}_{\rm eff}$ for the scalar field in a constant 
electric field, which is related to the probability 
of the system to remain in the vacuum 
state for all time, is given by~\cite{paddy91}
\begin{equation}
{\rm Im L}_{\rm eff} = \sum_{n=1}^{\infty} \frac{1}{2}\frac{(qE_0)^2}{(2\pi)^3}\frac{(-1)^{n+1}}{n^2}
\exp\left(-\frac{\pi m^2}{qE_0} n\right) 
\label{eqn:efflag}
\end{equation} 
where $m$ is the mass, $q$ is the charge and $E_0$ 
is the magnitude of the electric field. 
We will derive the above expression for 
${\rm Im L}_{\rm eff}$ using the general arguments 
given in section~(\ref{subsec:semiQM}).   

Consider the Hamilton-Jacobi equation for the motion 
of a  particle in an electromagnetic field in $(1+1)$ 
dimensions.
\begin{equation} 
\frac{1}{2}\left({\partial F \over \partial t} + qA^t 
\right)^2 - \frac{1}{2}\left({\partial F \over \partial x} 
- qA^x \right)^2 - \frac{1}{2} m^2
\; = \; 0 
\label{eqn:cpaths1}
\end{equation}
where $F$ is the action and $A^\mu = (A^t, A^x,0,0)$ 
is the four vector potential. 
We have neglected the dependence of $F$ on 
the $y$ and $z$ co-ordinates since it does not
change the results.  
In a time-dependent gauge, given by Eqn.~(\ref{eqn:gauge1}), 
the action $F$ can be easily solved for by using 
the ansatz $F = p_xx + f(t)$ to give,
\begin{equation}
F (t_1, x_1; t_0,x_0) \; = \; p_x(x_1 - x_0) 
\; \pm \; \int_{t_0}^{t_1} \! dt\sqrt{(p_x + qE_0t)^2 
+ m^2} 
\label{eqn:cpaths2t}
\end{equation}
where $p_x$ is the momentum of the particle in the 
$x$ direction.
The trajectory of the particle in the $(t,x)$ plane 
is the usual hyperbolic trajectory given by,
\begin{equation}
(t-t_i)^2 - (x-x_i)^2 = -\left({m \over qE_0}\right)^2 
\label{eqn:cpaths3}
\end{equation}
where $t_i$, $x_i$ are constants. 
For any fixed position $t$, there are thus two 
disjoint trajectories corresponding to motion 
in the two Rindler wedges. 
\par
Let us consider the tunnelling of a particle from 
one branch of the hyperbola to the other branch {\it and back} 
in the imaginary time co-ordinate $t$ (see Fig.1). 
This means that the particle comes back to the same 
spacetime point as it started from. 
Choosing the positive sign in  
Eqn.~(\ref{eqn:cpaths2t}) (this choice give a 
tunnelling probability that is exponentially damped), 
we have,
\begin{eqnarray}
F(t_0, x_0; t_0,x_0) &=& \oint \! dt
\sqrt{(p_x + qE_0t)^2 + m^2} = 
{m^2 \over qE_0} \oint du \sqrt{1 + u^2} 
\nonumber \\
&=& i{m^2\over qE_0}\oint d\tau \sqrt{1-\tau^2} = 
i{m^2\over qE_0}\int_{0}^{2\pi} d\theta 
\cos^2(\theta) \nonumber \\
&=& {i\pi m^2  \over qE_0} 
\label{eqn:cpaths4}
\end{eqnarray}
where we have made the following changes of variable 
$(p_x + qE_0 t)/m = u = i\tau$ and $\tau = \sin(\theta)$. 
The expression for $\exp(iF)$ 
from the above equation is seen to be exactly the same as 
the  exponential term in Eqn.~(\ref{eqn:efflag}) for $n=1$.
The same argument can be repeated for the particle tunnelling 
$n$ times to and fro to give 
\begin{equation}
F_n(t_0, x_0; t_0,x_0) = i{m^2\over qE_0}
\int_{0}^{2n\pi} d\theta \cos^2(\theta) = 
{i\pi m^2 \over qE_0}n . 
\label{eqn:cpaths4a}
\end{equation}
Again, the quantity $\exp(iF_n)$ is seen to 
match with the exponential part of the $n$th 
term in  Eqn.~(\ref{eqn:efflag}). 
Therefore, the imaginary part of the total effective 
lagrangian can be written down immediately as
\begin{equation}
{\rm Im L}_{\rm eff} = \sum_{n = 1}^{\infty} 
({\rm prefactor})\, \exp\left(-{\pi m^2 \over qE_0}n\right) 
\label{eqn:cpaths5}
\end{equation}
where the prefactor can only be calculated 
using the exact kernel. 
However, the dependence of the prefactor on $n$ and the phase 
factor $(-1)^{n+1}$ present in Eqn.~(\ref{eqn:efflag}) can 
be deduced using the following arguments. 
\par
The formal expression of the path integral kernel 
for the above electric field problem, in the time 
dependent gauge, is given by~\cite{paddy91} 
\begin{equation}
K(a,b;s) = \langle a\left| e^{ish} \right|b \rangle 
\label{eqn:cpaths6}
\end{equation}
where $K(a,b;s)$ is the kernel for the particle to 
propagate between the spacetime points $a=(x^0, 
{\bf x})$ and $b=(y^0, {\bf y})$ in a proper time $s$ 
and $h$ is the hamiltonian given by
\begin{equation}
h = \frac{1}{2} (i\partial_i - qA_i)(i\partial^i - qA^i) - 
\frac{1}{2}m^2 
\label{eqn:cpaths7}
\end{equation}
where $A^i$ is the four vector potential given 
in Eqn.~(\ref{eqn:gauge1}) and $q$ and $m$ are the 
charge and mass of the particle respectively. 
Going over to momentum co-ordinates and considering 
the coincidence limit ${\bf x} = {\bf y}$, the kernel 
can be written in the form,
\begin{equation}
K(x^0,y^0; {\bf x},{\bf x}; s) = -\frac{i}
{2 (2\pi)^2} \int_{-\infty}^{\infty}
 \! \frac{dp_x}{s} {\cal G}(x^0, y^0; s) 
\label{eqn:cpaths8}
\end{equation}
where ${\cal G}(x^0, y^0; s)$ is given by
\begin{equation}
{\cal G}(x^0, y^0; s) = \langle x^0 \left| e^{isH} 
\right| y^0\rangle 
\label{eqn:cpaths9}
\end{equation}
and $H$ is the hamiltonian 
\begin{equation}
H = -\frac{1}{2}\left( \frac{\partial^2}{\partial t^2} + 
(p_x + qE_0 t)^2 + m^2 \right) = -\frac{1}{2}
\left( \frac{\partial^2}{\partial \rho^2} + 
q^2E_0^2\rho^2 + m^2 \right) 
\label{eqn:cpaths10}
\end{equation}
with $\rho = (t + p_x/qE_0)$. In the expression for 
the kernel in momentum co-ordinates, we have integrated 
over the transverse momentum variables $p_y$ and $p_z$. 
The above hamiltonian is that of an inverted harmonic oscillator.  
(Since all reference to $p_y$ and $p_z$ have disappeared 
in $H$, the dependence of $F$ on the $y$ and 
$z$ co-ordinates was neglected when writing down 
the expression for the Hamilton-Jacobi 
equation in Eqn.~(\ref{eqn:cpaths1}).)  
The expression for the effective lagrangian is then 
given by
\begin{equation}
L_{\rm eff} = -i\int_0^{\infty}\! \frac{ds}{s} 
K(x^0,x^0; {\bf x},{\bf x}; s) =  -\frac{1}{2 (2\pi)^2}
\int_0^{\infty}\! \frac{ds}{s^2} 
\int_{-\infty}^{\infty} \! dp_x {\cal G}(x^0, x^0; s) 
\label{eqn:cpaths11}
\end{equation}
We would like to evaluate the propagator 
${\cal G}(x^0, x^0; s)$ for a tunnelling situation 
where the particle tunnels from the 
point $x^0$ and back in loops.  
Since the path integral is not well defined for 
closed loops, it will have to be evaluated in some 
approximate limiting procedure which is outlined below. 

Since the tunnelling potential is that of an inverted 
oscillator, we can use all the results 
of section~(\ref{subsubsec:potential1}) with 
the semi-classical wave functions given in Eqns.~(\ref{eqn:electpot2a},\ref{eqn:electpot2b}). 
We would like to first account for the 
factor $(-1)^n$ that arises when a particle tunnels 
from one side of the barrier to the other and back. 
 Consider an incident wave to the right of the barrier 
and impinging on it. 
Using the method of complex paths, we rotate this wave 
in the lower complex plane (this is the only route 
possible for the same reason as that given when rotating 
a right moving travelling wave in the upper complex plane) 
to obtain a wave again incident on the barrier with a 
(energy independent) phase factor $\exp(i\pi/2)$ being 
picked up (other factors dependent on the energy are 
also picked up but these are not important here). 
Since this wave is moving in the wrong direction, 
we assume that the particle that has tunnelled through 
has the same amplitude as the rotated wave but is moving 
away from the barrier.  
This just involves changing the sign of the argument of 
the exponential in the expression for the rotated wave. 
Rotating this left moving wave again in the upper complex 
plane now, the final wave obtained is a right moving wave with 
another extra phase factor of $\exp(i\pi/2)$ being picked up. 
The total phase change with respect to the 
incident wave is thus  $\exp(i\pi)$. 
Since this phase factor is independent of the 
energy, the propagator for the tunnelling process too, 
after one such rotation, will pick up 
a phase of $\exp(i\pi)$.  
Similarly, for $n$ rotations, $n$ taking the values 
$1,2,3,\ldots$, the phase acquired will be 
$\exp(in\pi) = (-1)^n$. 
\par
Therefore, the propagator ${\cal G}$ for $n$ loops, 
${\cal G}_n(x^0, x^0; s)$, can be written as
\begin{equation}
{\cal G}_n(x^0, x^0; s) = N(p_x, m, E) e^{in\pi} 
e^{iS_n(x^0, x^0; s)} 
\label{eqn:cpaths13}
\end{equation}
where $S_n(x^0, x^0; s)$ is the classical action for 
the tunnelling process and $N$ is the prefactor 
that arises after evaluating the ``sum over paths''. 
This prefactor is not expected to depend on the proper 
time $s$ since the tunnelling process takes place 
instantaneously or on the number of rotations $n$.  
So the only quantities it may depend on are 
$p_x$, $m$ and $E$.
Though the form of $N$ cannot be determined, we
can obtain the constraint on $N$ so as to give
the correct result thereby showing the existence
of such a factor.  
The action for the tunnelling process can be determined
by solving the Hamilton-Jacobi equation 
\begin{equation} 
\frac{1}{2}\left({\partial S \over \partial t} + qA^t 
\right)^2 - \frac{1}{2}\left({\partial S \over \partial x} 
- qA^x \right)^2 - \frac{1}{2} m^2 + 
\frac{\partial S}{\partial s} = 0 .
\label{eqn:cpaths14}
\end{equation}
The solution to the above equation is,
\begin{equation}
S  = -Es + p_xx \pm \int \! dt\sqrt{(p_x + qE_0t)^2 + 
(m^2 + 2E)} 
\label{eqn:cpaths14a} 
\end{equation}
Choosing the positive sign and setting $p_x + qE_0t = 
i\sqrt{m^2 + 2E}\sin(\theta)$, we obtain,
\begin{equation}
S  = -Es + p_xx \pm i\frac{m^2 + 2E}{2qE_0}
\int\! d\theta (1 + \cos(2\theta)) 
\label{eqn:cpaths15}
\end{equation}
For {\it closed} paths, with $\theta$ taking the 
values from $0$ to $2n\pi$, the above action can be written as,
\begin{equation}
S_n  = -Es \pm i\frac{m^2 + 2E}{2qE_0}\theta 
\label{eqn:cpaths16}
\end{equation}
We have thrown away the $p_x x$ term while retaining the 
$-Es$ term since the dependence on the $x$ 
co-ordinate is really irrelevant.  
Defining a new variable $\bar{\theta} = i\theta/2qE_0$ 
and rescaling $s=\alpha s'$, one obtains,
\begin{equation}
S_n  = -E\alpha s' + (m^2 + 2E)\bar{\theta} .
\label{eqn:cpaths17}
\end{equation}
Choosing $\alpha$ appropriately, $S_n$ can 
be cast into a form that matches the action for a 
fictitious free particle in $(1+1)$ dimensions with 
``energy'' $\alpha E$ and ``momentum'' $(m^2 + 2E)$ 
satisfying the classical energy-momentum relation,
\begin{equation}
\alpha E = \frac{1}{2}(m^2 + 2E)^2 
\label{eqn:cpaths18}
\end{equation}
where the particle's ``mass'' is set to unity for convenience.  
The above equation determines the quantity $\alpha$ 
and $S_n$ can be written in the form
\begin{equation}
S_n (\bar{\theta}_2, \bar{\theta}_1;s) = {(\bar{\theta}_2 -  
\bar{\theta}_1)^2 \over 2 s'} = {\alpha (\bar{\theta}_2 -  
\bar{\theta}_1)^2 \over 2 s} 
\label{eqn:cpaths19}
\end{equation}
where $\bar{\theta}_1$ and $\bar{\theta}_2$ are the 
initial and final states of the free particle with 
$s'$ being the proper time taken. (Note that 
$(\bar{\theta}_2 -  \bar{\theta}_1) = 2in\pi/2qE_0$.)  
Substituting this into the expression for 
${\cal G}_n(x^0, x^0; s)$ in Eqn.~(\ref{eqn:cpaths13}) 
and evaluating only the integral over $s$ in the 
expression for the effective lagrangian in 
Eqn.~(\ref{eqn:cpaths11}) {\it without} 
taking the limits, we obtain
\begin{eqnarray}
\int\! \frac{ds}{s^2}{\cal G}_n(x^0, x^0; s) &=& 
N(p_x, m, E) e^{in\pi} \int \! \frac{ds}{s^2} 
\exp\left({i\alpha (\bar{\theta}_2 -  
\bar{\theta}_1)^2 \over 2 s}\right) \nonumber \\
&=& -N(p_x, m, E) e^{in\pi} \frac{2}{i\alpha 
(\bar{\theta}_2 -  \bar{\theta}_1)^2} 
\exp\left({iS_n(\bar{\theta}_2, \bar{\theta}_1;s)}\right) .
\label{eqn:cpaths20}
\end{eqnarray}
Notice that the prefactor to the exponential term has no 
dependence on the proper time $s$. 
Now, we use the form for the action given by 
Eqn.~(\ref{eqn:cpaths16}) and taking the limits 
for $s$ from $0$ to $\infty$, the effective 
lagrangian for $n$ loops, $L_{\rm eff} (n)$, 
can be written in the form, 
\begin{equation}
L_{\rm eff} (n) = \frac{i}{2} \frac{(qE_0)^2}{(2\pi)^3} 
\frac{(-1)^{n+1}}{n^2} \exp\left(-\frac{\pi m^2}
{qE_0} n\right) \frac{8}{\alpha} 
\exp\left(-\frac{2\pi E}{qE_0} n\right) 
\int_{-\infty}^{\infty} \! \frac{dp_x}{2\pi} N(p_x, m, E)  
\label{eqn:cpaths21}
\end{equation}
where we have set $(\bar{\theta}_2 -  \bar{\theta}_1) 
= 2in\pi/2qE_0$. 
Taking the limit $E\to 0$, so as to obtain
the correct result, and using the 
expression for $\alpha$ in Eqn.~(\ref{eqn:cpaths18}), 
we find that $N$ must satisfy the relation
\begin{equation}
\lim_{E\to 0} \frac{16E}{(m^2 + 2E)^2} \,
\int_{-\infty}^{\infty} \! \frac{dp_x}{2\pi} 
N(p_x, m, E)  = 1 
\label{eqn:cpaths22}
\end{equation}
so that the imaginary part of the effective 
lagrangian for $n$ loops, $L_{\rm eff}(n)$, 
matches the $n$th term in Eqn.~(\ref{eqn:efflag}). 
Therefore, in this manner, the contributions to 
the imaginary part of the effective lagrangian 
for the uniform electric field can be thought of 
as arising from the tunnelling of particles 
between the two Rindler sectors.

\section{Conclusions}
In conclusion, we see that particle production can be obtained 
in Schwarchild-like spacetimes in the standard co-ordinate 
systems without requiring the maximally extended manifold.  
The method of complex paths used in ordinary quantum mechanics 
is modified appropriately to produce a prescription that 
regularises the singularity in the action functional and 
Hawking radiation is recovered as a consequence.  
In the case of the electric field, particle production in 
different gauges has been described using the tunnelling 
description which gives a correspondence between the 
transmission and reflection coefficients and the 
standard Bogolubov coefficients. 
The interesting feature of the mixed gauges that were 
considered was that the mode functions could be 
combinations of elementary functions for certain values 
of the gauge parameters.         
The method of complex paths also gives a 
simple interpretation of particle production in an 
electric field as arising due to tunnelling between 
the two disjoint sectors of the Rindler spacetime. 
Though we have only given a heuristic argument in 
this paper, we will explore this issue further 
in a future publication.

\section*{Acknowledgements}
\noindent
KS is being supported by the Senior Research Fellowships 
of the Council of Scientific and Industrial Research, India.

\appendix

\section{Facets of Tunnelling}
\label{sec:facets}
In this section we briefly review the basic concepts of 
semi-classical quantum mechanics in one dimension and 
formally describe the tunnelling process.  
We then apply the formalism to two potentials, namely, 
$V_1(x) = -x^2$ and $V_2(x) = -1/x^2$, and calculate the 
transmission and reflection coefficients for both.  

\subsection{Semi-classical limit of Quantum Mechanics} 
\label{subsec:semiQM}
Consider a simple one dimensional quantum mechanical system with 
an arbitrary potential $V(x)$ where $x$ denotes the space 
variable (see Ref.~\cite{landau3} for details). 
To describe the transition of the system from one state to another, 
we first solve the corresponding classical equations of motion 
and determine the path of transition.  
This path is, in general, complex since many processes like 
tunneling through a potential barrier cannot occur classically.  
Therefore, the transition point $q_0$ where the system formally 
makes the transition is a complex number  determined by the 
classical conservation laws.  
Then, the action $S$ for the transition from some initial state 
$x_i$ to a final state $x_f$ given by
\begin{equation}
S(x_f, x_i) = S(x_f, q_0) + S(q_0, x_i)
\end{equation}
is calculated.  
Here, $S(q_0, x_i)$ is the action for the system to move from the 
initial state $x_i$ to the transition point $q_0$ while 
$S(x_f, q_0)$ is that to move from $q_0$ to $x_f$.  
The probability $P$ for the transition to occur is given by the formula 
\begin{equation}
P \sim \exp \left( -{2\over \hbar} {\rm Im}[S(x_f, q_0) + 
S(q_0, x_i)]\right) .
\label{eqn:semiclas1}
\end{equation}
The above formula is valid only when the argument of the 
exponential is large.  
Further, if the potential energy has singular points, these 
must also be considered as possible values for $q_0$.
If the position of the transition point is not unique, then it 
must be chosen so that the exponent in Eqn.~(\ref{eqn:semiclas1}) 
has the smallest absolute value but still must be large enough 
so that the above formula be valid.  

If the transition point $q_0$ is real, but lies in the classically 
inaccessible region, then the above formula gives the transmission 
coefficient for penetration through a potential barrier, while if 
the transition point is complex, it solves the problem finding 
the over the barrier reflection coefficient. 
The $\sim$ sign in the above formula is used since the coefficient 
in front of the exponential is not determined. 
This can be determined by calculating the exact semi-classical wave 
functions.  
Generally, it is desirable to find the ratios of two different 
transitions so that this coefficient does not matter. 

The physics of the tunnelling and the ``over the barrier'' 
reflection processes are very different.  
In the tunnelling process, the semi-classical analysis gives a 
transmission coefficient that is an exponentially small quantity 
with the corresponding reflection coefficient being unity.  
In contrast, in the ``over the barrier'' reflection process, 
 just the reverse is obtained.  
The transmission coefficient is unity while the reflection 
coefficient is an exponentially small quantity. 
Both these processes will be encountered when the electric 
field is studied in different gauges. 

We will now review the method of calculating the transmission and 
reflection coefficients for a typical quantum mechanical problem 
using the method of complex paths for a general potential $V(x)$.  

\subsubsection{Description of the method of complex paths} \label{subsubsec:tunnel}

Consider the motion of a particle of mass $m$ in a region 
characterised by the presence of a potential $V(x)$ in one 
space dimension. 
The problem is to calculate the transmission and reflection 
coefficients between two asymptotic regions labelled $L$ and 
$R$ where the semi-classical approximation to the exact wave 
function is valid.  
After identifying these regions and writing down the 
semi-classical wave functions, definite boundary conditions 
are imposed.  
The usual boundary conditions considered  are that in one 
region, say $L$,  the wavefunction is a superposition of an 
incident wave and a reflected wave while in the second region 
$R$, the wave function is just a transmitted wave.  
Then, a complex path (in the plane of the now complex 
variable $x$) is identified from $R$ to $L$ such that  
(a) all along the path the semiclassical ansatz is valid 
and (b) the reflected wave is exponentially greater than 
the incident wave at least in the latter part of the path 
near the region $L$.  
The transmitted wave is then moved along the path  to obtain 
the reflected wave and thus the amplitude of reflection 
is identified in terms of the transmission amplitude.  
Having done this, the normalisation condition is used {\it i.e.} 
the sum of the modulus square of the transmission and reflection 
amplitudes should equal unity, to determine the exact values 
of the transmission and reflection coefficients.   

For a given potential, the turning points $q_0$ (or transition 
points) are given by solving the equation
\begin{equation}
p(q_0) = \sqrt{2m(E - V(q_0))} = 0 \label{eqn:tpoint1}
\end{equation}
where $p(x)$ is the classical momentum of the particle and 
$E$ is the energy of the particle. 
In general, $q_0$ is complex.  
At these points, the semi-classical ansatz is not valid since 
the momentum is zero. 
Further, the potential can possess singularities. 
At these points too, the semi-classical approximation is invalid.  
Therefore the contour between the two regions should be chosen to 
be far away from such points. 
In general the contour will enclose them.  
Therefore, the relation between the transmission and reflection 
amplitudes is determined by taking into account the turning points 
and the singularities of the potential.     

The Schr\"odinger equation to determine the wave function $\psi$ 
of the particle is
\begin{equation}
-{\hbar^2 \over 2m} {d^2 \psi \over dx^2}= (E - V(x))\psi . \label{eqn:schrodinger}
\end{equation}
Referring to~\cite{landau3}, the  semi-classical wave 
function, in the classically allowed region where $E>V(x)$, 
is given by the formula
\begin{equation}
\psi = C_1 p^{-1/2} \exp\left( {i\over \hbar} 
\int \! p(x)dx \right) + C_2 p^{-1/2} \exp\left(-{i\over \hbar} 
\int \! p(x)dx \right) 
\label{eqn:tunnel1}
\end{equation}
while in the classically inaccessible regions of space where 
$E<V(x)$, the function $p(x)$ is purely imaginary and the wave 
function is now given by the relation
\begin{equation}
\psi = C_1 |p|^{-1/2} \exp\left( -{1\over \hbar} 
\int \! |p(x)|dx \right) \;+\; C_2 |p|^{-1/2} \exp\left({1\over \hbar} 
\int \! |p(x)|dx \right) .
\label{eqn:tunnel2}
\end{equation}
The condition on the potential for semi-classicality of the 
wave function to be valid is 
\begin{equation}
\left| {d \over dx}\left({\hbar \over p(x)}\right) \right| \ll 1, \label{eqn:cond1}
\end{equation}
or, in another form,
\begin{equation}
{m\hbar |F| \over |p|^3} \ll 1, \qquad F = -{dV \over dx}. 
\label{eqn:cond2}
\end{equation}
It ought to be noted that the accuracy of the semi-classical 
approximation is not such as to allow the superposition of 
exponentially small terms over exponentially large ones. 
Therefore, it is inapplicable in general to retain both terms 
in Eqn.~(\ref{eqn:tunnel2}).  
We will consider a  few cases of interest in this paper and  
refer the reader to \cite{landau3} for a exhaustive discussion 
along with suitable illustrative examples.

Consider the case in which the semi-classical 
condition~(\ref{eqn:cond2}) holds in the regions 
$x\to \pm\infty$. 
As $x\to -\infty$, the wave function is assumed to be a 
superposition of incident and reflected waves and is written 
in the form, 
\begin{equation}
\psi =  p^{-1/2} \exp\left( {i\over \hbar} 
\int \! p(x)dx \right) + C_2 p^{-1/2} \exp\left(-{i\over \hbar} 
\int \! p(x)dx \right) 
\label{eqn:tunnel10}
\end{equation}
where the incident wave has unit amplitude while the 
reflected wave has amplitude given by $C_2$. As $x\to +\infty$, 
the wave function is assumed to be a  right moving travelling wave, 
\begin{equation}
\psi = {C_3 \over \sqrt{p}} \exp \left( {i \over \hbar} 
\int \! p dx  \right)  
\label{eqn:tunnel11}
\end{equation}
The method of complex paths can now be  applied on the function~(\ref{eqn:tunnel11}).  The contour is chosen either 
in the upper or lower complex plane such that the reflected 
wave is always exponentially greater than the incident wave 
along that part of the path near the region $x \to -\infty$.  
If this is satisfied along one of the contours then $C_2$ 
is determined in terms of $C_3$. To carry out the above 
procedure however, the exact semi-classical wave functions 
as $x\to \pm \infty$ have to be determined. 
This will be done explicitly later for the relevant cases. 

A different case arises when the semi-classical ansatz holds 
in the vicinity of the origin $x=0$ rather than at $x=\pm \infty$. 
The boundary conditions assumed in this case are the same 
as above with the condition $x \to \infty$ replaced by 
$x \to 0^+$ and $x \to -\infty$ by $x \to 0^-$. 
Here, the required contour is about the origin and is chosen 
to be small.  
But it must still be large enough so that the reflected 
wave is much larger than the incident wave along the latter 
part of the contour near the region $x < 0$.  

In the above cases the method of complex paths gives the 
exact transmission and reflection amplitudes.  
But, in certain cases it is enough to assume that the 
transmission and incident amplitudes are equal to unity 
while the reflection amplitude is exponentialy damped and 
consequently very small. 
Here, the ``over the barrier'' reflection coefficient for 
energies large enough so that the reflection coefficient is 
exponentially small, has to be determined. 
In this case, the condition $E>V(x)$ is always satisfied.  
Therefore, the transition point $q_0$ at which the particle 
reverses its direction is the complex root of the equation 
$V(q_0) = E$.  
Let $q_0$ lie in the upper complex plane for definiteness. 
Now, the amplitudes of the incident and transmitted waves 
are equal (both are set to unity within exponential accuracy).  
To calculate the reflection coefficient, the relation between 
the wave functions far to the right of the barrier and far 
to the left of the barrier must be determined.  
The transmitted wave can be written in the form
\begin{equation}
\psi_T = {1 \over \sqrt{p}} \exp\left( {i \over \hbar} 
\int_{x_1}^{x} \! p dx \right) 
\label{eqn:aboveb1}
\end{equation}
where $x_1$ is any point on the real axis.  
We follow the variation of $\psi_T$ along a path $C$ in the upper 
complex plane which encloses the turning point $q_0$.   
The latter part of this path must lie far enough to the left 
of $q_0$ so that the error in the semi-classical incident wave 
is less than the required small reflected wave. 
Passage around $q_0$ only causes a change in the sign of the 
root $\sqrt{E -V(x)}$ and after returning to the real axis, 
the function $\psi_T$ becomes the reflected wave $\psi_R$. 
Going around a complex path in the lower complex plane 
converts $\psi_T$ into the incident wave.  
Since the amplitudes of the incident and transmitted waves may 
be regarded as equal, the required reflection coefficient 
is given by 
\begin{equation}
R = \left|{\psi_R \over \psi_T} \right|^2 = 
\exp\left( - {2 \over \hbar}{\rm Im}\int_{C} \! pdx \right) 
\label{eqn:aboveb2}
\end{equation}
Now we can deform the contour in any way provided it still 
encloses the point $q_0$.  
In particular, the contour can be deformed to go from 
$x_1$ to $q_0$ and back.  
This gives
\begin{equation}
R = \exp\left( - {4 \over \hbar}{\rm Im}\int_{x_1}^{q_0} \! pdx 
\right) . 
\label{eqn:aboveb3}
\end{equation}
Since $p(x)$ is real everywhere, the choice of $x_1$ on the 
real axis is immaterial. 
The above formula determines the above the barrier reflection 
coefficient. 
It must be emphasised that to apply the above formula the 
exponent must be large so that $1-R$ is very nearly equal to unity. 

Finally consider a situation where the amplitudes of the 
reflection and incident wave are equal.  
The transmission coefficient is now an exponentially small 
quantity. 
This case corresponds to the standard tunnelling process. 
The problem is characterised by the presence of real turning 
points between which lies the classically forbidden region 
where the energy $E < V(x)$.  
For definiteness,  let there be two real turning points 
labelled $q_{-}$ and $q_{+}$. The potential $V(x)$, in the 
immediate vicinity of the turning points of the barrier, 
is assumed to be of the form
\begin{equation}
E- V(x) \approx F_0 (x-q_{\pm}), \qquad F_0 = -\left. 
{dV \over dx}\right|_{x = q_\pm}
\label{eqn:cond3}
\end{equation}
This assumption is equivalent to saying that the particle, 
near the turning points, moves in a homogeneous field.  
With this assumption, the amplitude of transmission $C_3$ 
is given by (refer to Ref.~\cite{landau3} page $181$)
\begin{equation}
C_3 =  \exp\left( -{1\over \hbar} \int_{q_{-}}^{q_{+}} \! 
|p(x)|dx  \right) 
\label{eqn:barrier1}
\end{equation} 
The transmission coefficient is then given by $|C_3|^2$. 
The above formula holds only when the exponent is large. 
In the derivation above, we have assumed that the 
semi-classical condition holds across the entire barrier 
except in the immediate vicinity of the turning points. 
In general, however, the semi-classical condition need not 
hold over the entire extent of the barrier. 
The potential, for example, could drop steeply enough so 
that Eqn.~(\ref{eqn:cond3}) is not valid. 
In these cases, the exact semi-classical equations have to 
be determined before applying the method of complex paths. 
The cases encountered in this paper all satisfy 
Eqn.~(\ref{eqn:cond3}). 

We now apply the above results to two potentials.  
The first is the well known inverted harmonic oscillator 
potential $V_1(x) = -g_1x^2$ with $g_1>0$ while the other 
is $V_2(x) = -g_2/x^2$ with $g_2 > 0$.  
The first potential arises when the propagation of a scalar 
field in a constant electric field background is studied.  
The second potential arises when the propagation of a scalar 
field in Schwarzchild-like spacetimes is considered in the 
vicinity of the horizon. 
 
\subsubsection{\it Application to the potential $V_1(x) = -g_1 x^2$} \label{subsubsec:potential1}
Consider the potential given by 
\begin{equation}
V_1(x) = -g_1 x^2 \label{eqn:pot1}
\end{equation}
where $g_1>0$ is a constant.  
This potential is the inverted harmonic oscillator potential 
and is discussed extensively in many places (see for 
example \cite{bandd82}, \cite{landau3}). 
We will follow the semi-classical treatment given in 
Ref.~\cite{landau3} and briefly review the calculation of 
the reflection and transmission coefficients for both the 
tunneling and over the barrier reflection cases.

The semi-classicality condition~(\ref{eqn:cond2}) for 
the above potential is, 
\begin{equation}
\left| {\hbar g_1 x \over \sqrt{2m} [E_1 + g_1 x^2]^{3/2}} 
\right| \ll 1 
\label{eqn:pot1cond}
\end{equation}
where $m$ is the mass of the particle and $E_1$ is its energy.  
The above condition definitely holds for large enough $|x|$ and 
for any value of $E_1$, either positive or negative. 
Therefore the motion  of a particle moving under such 
a potential is semi-classical for large enough $|x|$ and 
hence holds as $x \to \pm \infty$.   

Since the motion is semi-classical for large $|x|$, we can 
expand the momentum $p(x)$ as,
\begin{equation}
p(x) = \sqrt{2m\left( E_1 + g_1x^2 \right) } \approx  
\sqrt{2mg_1}\left( x + {E_1 \over 2g_1 x} \right) 
\label{eqn:electpot1} 
\end{equation}
Using Eqn.~(\ref{eqn:electpot1}), the  semi-classical wave 
functions can be written as follows with the following 
boundary conditions. 
As $x\to \infty$, we assume that the wave function is a 
right moving travelling wave $\psi_R$ while as $x\to -\infty$, 
it is a superposition of an incident wave of unit amplitude 
and a reflected wave given by $\psi_L$.  
Therefore, we have, 
\begin{eqnarray}
\psi_R = C_3 \xi^{i\varepsilon_1 - {1\over 2}} e^{i\xi^2/2} 
\qquad \qquad \qquad \qquad \qquad \; && 
\qquad \qquad (\xi \to +\infty) \label{eqn:electpot2a}  \\
\psi_L =  (-\xi)^{- i\varepsilon_1 - 
{1\over 2}} e^{- i\xi^2/2} + 
C_2 (-\xi)^{i\varepsilon_1 - {1\over 2}} e^{i\xi^2/2} && 
\qquad \qquad (\xi \to -\infty) 
\label{eqn:electpot2b}
\end{eqnarray}
where we have made the definitions
\begin{equation}
\xi = \left({2mg_1 \over \hbar^2}\right)^{\! 1/4} x \qquad 
; \qquad \varepsilon_1 =  {1\over \hbar}\sqrt{m \over 2g_1} E_1 \label{eqn:electpot3}
\end{equation}
Following the variation of Eqn.~(\ref{eqn:electpot2a}) around 
a semi-circle of large radius $\rho$ in the {\it upper} 
half plane of the now complex variable $\xi$, we obtain 
\begin{equation}
C_2 = -iC_3 \exp\left( -\pi\varepsilon_1 \right) 
\label{eqn:electpot4}
\end{equation}
The conservation of particles is expressed by the condition 
that, 
\begin{equation}
|C_3|^2 + |C_2|^2 = 1 \label{eqn:pconservation}
\end{equation}
From Eqn.~(\ref{eqn:electpot4}) and 
Eqn.~(\ref{eqn:pconservation}), the transmission coefficient 
is, 
\begin{equation}
T = |C_3|^2 = {1 \over 1 + e^{-2\pi\varepsilon_1} } = 
{1 \over 1 + e^{-{1 \over \hbar} \pi \sqrt{2m/g_1}E_1} } 
\label{eqn:electpot5}
\end{equation}
while the reflection coefficient is 
\begin{equation}
R = |C_2|^2 =  {e^{-{1 \over \hbar} \pi \sqrt{2m/g_1}E_1} 
\over 1 + e^{-{1 \over \hbar} \pi \sqrt{2m/g_1}E_1} } 
\label{eqn:electpot6}
\end{equation}
Note that the passage through the {\it lower} half 
complex plane to determine $C_2$ is unsuitable since on the 
part of the path $\, -\pi < \phi< -\pi/2 \,$, where $\phi$ 
is the phase of the complex variable $\xi$, the incident wave 
(first term in Eqn.~(\ref{eqn:electpot2b})) is exponentially 
large compared with the reflected wave. 
The above formula holds for all energies $E_1$.  
This is because, even for negative energies, the 
semi-classical wave functions given in Eqns.~(\ref{eqn:electpot2a},\ref{eqn:electpot2b}) are 
exactly the same with the boundary conditions being fully 
satisfied.   

If $E_1$ is large and negative, Eqn.~(\ref{eqn:electpot5}) 
gives $T \approx e^{-\pi\sqrt{2m/g_1}|E_1|/ \hbar}$ and 
thus $R\sim 1$. 
This is in accordance with the formula in Eqn.~(\ref{eqn:barrier1}). 
To apply Eqn.~(\ref{eqn:barrier1}) it is necessary to calculate 
the turning points first. 
The real turning points are $q_0 = -\sqrt{|E_1|/g_1}$ and 
$q_1 = \sqrt{|E_1|/g_1}$.  
Therefore,
\begin{eqnarray}
C_3 &=&  \exp\left( -{1\over \hbar} 
\int_{q_0}^{q_1} \! |p(x)|dx  \right) \nonumber \\
&=&  \exp\left(-{1\over \hbar}\sqrt{2mg_1}\int_{q_0}^{q_1} \! 
\left|\sqrt{x^2 - {|E_1|\over g_1}}\right| dx \right) 
\nonumber \\
&=& \exp\left(-{1\over 2\hbar}\pi \sqrt{2m/g_1}|E_1| \right) \label{eqn:electpot61}
\end{eqnarray} 
This gives the same answer. 

We can calculate the over the barrier reflection coefficient 
using Eqn.~(\ref{eqn:aboveb3}) for $E_1$ large and {\it positive}.  
The turning points now are given by solving the equation 
$p(q_0) = 0$ with the condition that $E_1>V_1(x)$ always.  
Since $E_1 > 0$, the turning points are  $q_0 = \pm i\sqrt{E_1/g_1}$. 
Choosing the positive sign for $q_0$ and setting $x_1 = 0$, 
the integral in Eqn.~(\ref{eqn:aboveb3}) is evaluated as follows:  
\begin{eqnarray}
\int_{0}^{q_0}\! p(x) dx &=& \sqrt{2mg_1} 
\int_{0}^{q_0}\!\sqrt{E_1/g_1 + x^2} \nonumber \\
&=& i\sqrt{2m/g_1}E_1\int_{0}^{1}\!\sqrt{1 - y^2} = 
{1 \over 4}i\pi \sqrt{2m/g_1}E_1
\label{eqn:electpot7}
\end{eqnarray}
Therefore, 
\begin{equation}
R = \exp\left( - {1 \over \hbar}\pi \sqrt{2m/g_1}E_1 \right) \label{eqn:electpot8} 
\end{equation}
The above formula can also be obtained directly from 
Eqn.~(\ref{eqn:electpot6}) by neglecting the exponential 
term compared to unity which means that the energy has 
to be large.  

\subsubsection{\it Application to the potential $V_2(x) 
= -g_2/x^2$} 
\label{subsubsec:potential2}
Consider the potential given by 
\begin{equation}
V_1(x) = -{g_2 \over x^2} \label{eqn:pot2}
\end{equation}
where $g_2$ is a positive constant. 
The potential has a singularity at the origin.  
This potential arises when the effective Schr\"odinger 
equation is calculated for Schwarzchild-like spacetimes. 
This aspect will be dealt with in later sections.  

The semi-classical condition~(\ref{eqn:cond2}) for this 
potential takes the form
\begin{equation}
\left|{\hbar g_2 \over \sqrt{2m}}
{1 \over [E_2 x^2 + g_2]^{3/2} }\right| \ll 1 
\label{eqn:pot2cond}
\end{equation} 
where $E_2$ is the energy. 
It is clear that the above relation holds for large $|x|$. 
It also holds for small $|x|$ if $\sqrt{2 m g_2} \gg \hbar$. 
Notice that the quasi-classicality condition for small $|x|$ 
is independent of the sign and magnitude of the energy $E_2$. 
For this potential, we will be concerned only with the small 
$|x|$ behaviour in contrast with the potential $V_1$.  
Since the motion is semi-classical for small $|x|$, 
we expand the momentum $p(x)$ as
\begin{equation}
p(x) = \sqrt{ 2m \left(E_2 + {g_2\over x^2} \right) } 
\approx {\sqrt{2mg_2} \over x} + \sqrt{m \over 2g_2}E_2 x 
\label{eqn:bhpot1}
\end{equation}
Notice the similarity between Eqn.~(\ref{eqn:electpot1}) 
and Eqn.~(\ref{eqn:bhpot1}). 

We will calculate the over the barrier reflection 
coefficient with $E_2 > 0$ and small, which will be 
of interest later. Using the expansion in 
Eqn.~(\ref{eqn:bhpot1}), the semi-classical wave 
functions with the following boundary conditions, 
namely, that for $x>0$ the wave function is a right 
moving travelling wave while it is a superposition of 
an incident wave of unit amplitude and reflected wave 
for $x<0$,  are
\begin{eqnarray}
\psi_R = C_3 \xi^{ i\varepsilon_2 + {1\over 2}} 
e^{ i\xi^2/2} \qquad \qquad \qquad && \qquad \qquad 
(\xi > 0) \label{eqn:bhpot2a} \\ 
\psi_L =  (-\xi)^{- i\varepsilon_2 + {1\over 2}} 
e^{- i\xi^2/2} + C_2 (-\xi)^{ i\varepsilon_2 + {1\over 2}} 
e^{i\xi^2/2} && \qquad \qquad (\xi < 0) 
\label{eqn:bhpot2b}
\end{eqnarray}
where we have made the definitions
\begin{equation}
\xi = \left({mE_2^2 \over 2g_2}\right)^{1/4}\! x \qquad 
; \qquad \varepsilon_2 = {\sqrt{2mg_2} \over \hbar} 
\label{eqn:bhpot3}
\end{equation}
Following the variation of Eqn.~(\ref{eqn:bhpot2a}) around 
an small semi-circle of radius $\rho < \sqrt{g_2/|E_2|}$ 
(in contrast to the potential $V_1$ where the radius 
$\rho$ was large) in the {\it upper } half complex plane, 
we obtain,
\begin{equation}
 C_2 = C_3 \exp\left( -\pi\varepsilon_2 +{i\pi \over 2} 
\right) 
\label{eqn:bhpot4}
\end{equation}
Setting $T = |C_3|^2 = 1$ and $R = |C_2|^2$, we 
finally obtain, 
\begin{equation}
R = T e^{-2\pi\varepsilon_2} = T 
e^{-{1 \over \hbar}2\pi\sqrt{2mg_2}} 
\label{eqn:bhpot5}
\end{equation}
Using the normalisation condition $R + T = 1$, we 
obtain,
\begin{equation}
T = { 1 \over 1 + e^{-{1 \over \hbar}2 \pi\sqrt{2mg_2}} } 
\quad {\rm  and } \quad R =  { e^{-{1\over \hbar}2 \pi\sqrt{2mg_2}} 
\over 1 + e^{-{ 1\over \hbar}2 \pi\sqrt{2mg_2}} } 
\label{eqn:bhpot51}
\end{equation}
Notice that the above result is independent of the energy 
$E_2$ and hence holds for $E_2 < 0$ too. For small $|x|$, 
the lack of dependence on $E_2$ is not too surprising since 
the contour is such that it is not too close to the real 
turning points $q_0 = \pm \sqrt{g_2/|E_2|}$.  
Anyway, when $E_2\sim 0^{+}$, $\rho$ is ``large'' and 
therefore the contour is chosen to lie in the upper complex 
plane for the same reason as given in the analysis of the 
potential $V_1$ in the previous section. 

We will derive the result in Eqn.~(\ref{eqn:bhpot5}) 
using Eqn.~(\ref{eqn:aboveb2}).  
The complex turning points $q_0$ are the roots of the equation 
$E_2 = -g_2/q_0^2$ where $E_2>0$ and therefore, the turning 
points are $q_0 = \pm i\sqrt{g_2/E_2} = \pm i p_0$.  
Hence, we have to evaluate the integral,
\begin{equation}
\int_C \! pdx = \sqrt{2mE_2}\int_C \! 
\sqrt{1 + {p_0^2 \over x^2}} dx 
\label{eqn:bhpot7}
\end{equation}
where the contour $C$ encircles the point $x = ip_0$ in 
the upper half complex plane.  
However, since there is a singularity at the origin, 
we cannot deform the contour as was done when deriving 
Eqn.~(\ref{eqn:aboveb3}).  
Therefore, as a means of regularisation, we modify the 
potential to 
\begin{equation}
V_{\rm mod}(x) = -{g_2 \over x^2 + \epsilon^2} 
\label{eqn:bhpot8}
\end{equation}
where the limit $\epsilon \to 0$ must be taken at the end 
of the calculation.   
The turning points for the modified potential are 
$x_{\rm mod} = \pm i\sqrt{\epsilon^2 + g_2/E_2}$ while the 
poles of the modified potential are at 
$x = \pm i\epsilon < x_{\rm mod}$. 
Even in this case, there is a singularity on the path of 
integration which contributes to the integral rather than 
the turning point.  
Therefore, integrating upto $+i\epsilon$ using the modified 
potential and back, we obtain,  
\begin{eqnarray}
\int_C \! pdx &=& \lim_{\epsilon \to 0} 
2\sqrt{2mE_2}\int_{0}^{i\epsilon}\! 
\sqrt{ 1 + {p_0^2 \over x^2 + \epsilon^2}} dx \nonumber \\
&=& \lim_{\epsilon \to 0} 2i\sqrt{2mE_2}\epsilon \int_{0}^{1} 
\! dy \sqrt{ 1 +  {p_0^2/\epsilon^2 \over 1 - y^2} } 
\nonumber \\
&\approx & 2i\sqrt{2mE_2}p_0 \int_o^1 \! {dy \over \sqrt{1-y^2}} 
\nonumber \\
&=& i\pi \sqrt{2mE_2}p_0 = i\pi \sqrt{2mg_2} 
\label{eqn:bhpot9}
\end{eqnarray}   
We therefore recover the result given in Eqn.~(\ref{eqn:bhpot5}). 
From the above calculation it is clear that, due to the 
singularity at the origin, the reflection coefficient has no 
contribution from the turning point at all.

\begin{figure}
\psfig{file=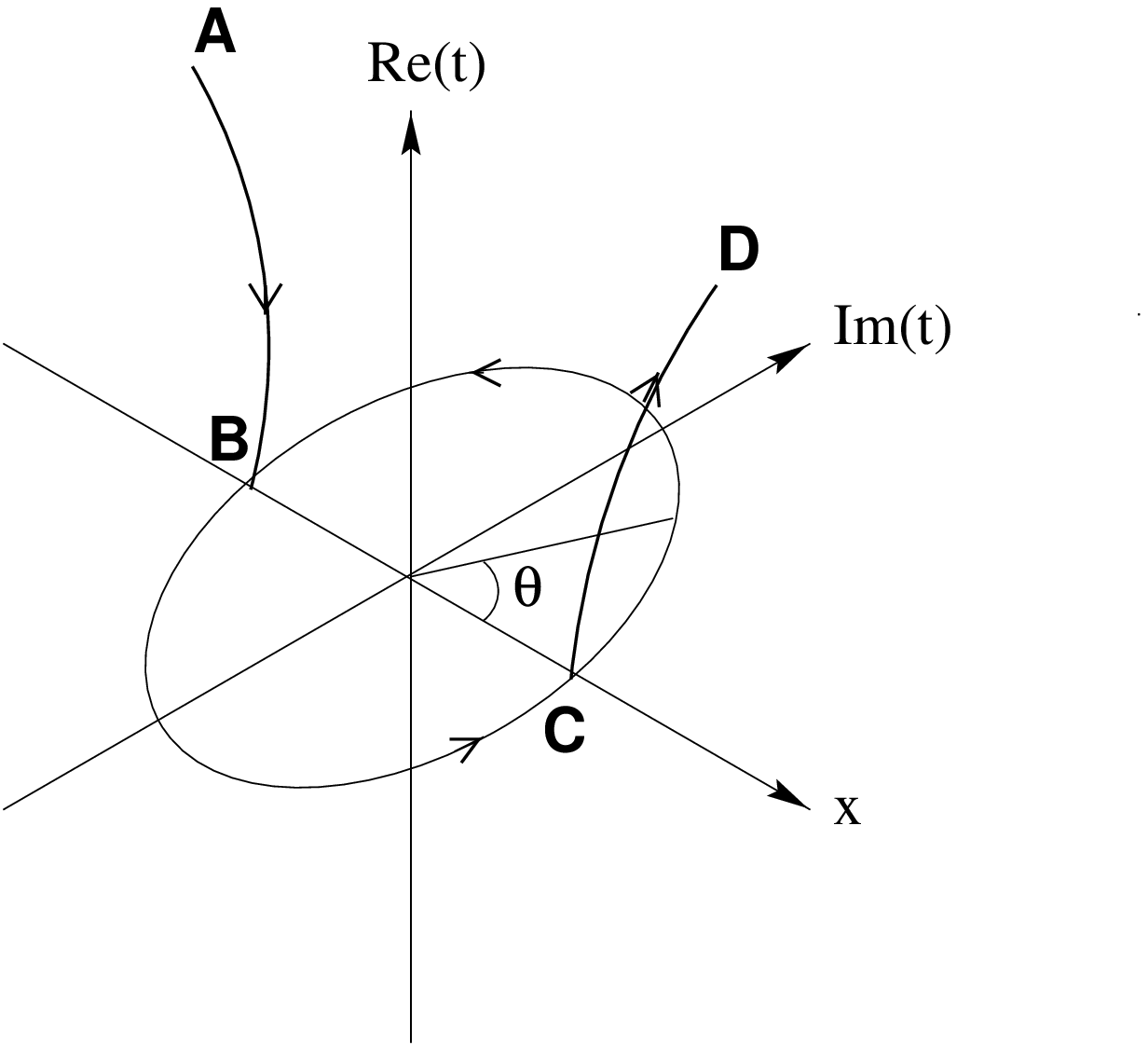}
\caption[Tunnelling in Imaginary time]
{$AB$ and $CD$ are the hyperbolic branches representing motion in the two disjoint Rindler sectors for a fixed $t$. The particle tunnells from $B$ to $C$ and back in a circular path along the imaginary time-space plane. If the particle makes $n$-loops around the circle, the imaginary part of the action
will be $n$ times the value for single loop.}
\end{figure}


\begin{thebibliography}{20}
\bibitem{schwinger}
J.~Schwinger, Phys.Rev., {\sl On Gauge Invariance 
and Vacuum Polarization},\ {\bf 82}, 664 (1951)
\bibitem{brout95}
R.~Brout, S.~Massar, R.~Parentani, P.~Spindel, 
Phys.\ Rept.\ {\bf 260} 329-454 (1995) 
and references cited therin.
\bibitem{paddy91}
T.~Pamanabhan, Pramana--J.\ Phys.\  {\bf 179}, 
37 (1991).
\bibitem{hawking76}
J.B.~Hartle and S.W.~Hawking, Phys.\ Rev.\ D. 
\ {\bf 189}, 13 (1976).
\bibitem{landau3}
L.~D.~Landau and E.~M.~Lifshitz, 
{\sl Quantum Mechanics (Non-relativistic Theory)}, 
Course of Theoretical Physics, Volume~2  
\  (Pergamon Press, New York, 1975).
\bibitem{stephens88}
C.~R.~Stephens, Ann. Phys. (N.Y.) {\bf 181}, 120 (1988).
\bibitem{bandd82}
N.~D.~Birrel and P.~c.~W.~Davies, 
{\sl Quantum Field in Curved Space}\ 
(Cambridge University Press, Cambridge, England, 1982). 
\bibitem{schulman81}
L.~S.~Schulman, {\sl Techniques and Applications of 
Path Integration} \ (John Wiley, New York, 1981).  
\bibitem{fulling89}
S.~A.~Fulling, {\sl Aspects of Quantum Field Theory 
in Curved Spacetime}
\ (Cambridge University Press, Cambridge, England, 1989).
\bibitem{sriram97}
L.~Sriramkumar, {\sl Quantum Fields in Non-trivial 
Backgrounds},\  Ph.D thesis, IUCAA (1997).
\bibitem{gandr80}
I.~S.~Gradshteyn and I.~M.~Ryzhik, {\sl Table of 
Integrals, Series and Products} \ (Academic, 
New York, 1980).
\bibitem{arfken85}
G.~Arfken, {\sl Mathematical Methods for Physicists} 
\ (Academic, New York, 1985).


\end{thebibliography}
\end{document}